\documentclass[letterpaper, twocolumn, superscriptaddress, showpacs, preprintnumbers, amsmath, amssymb]{revtex4}
\usepackage{graphicx}
\usepackage{dcolumn}
\usepackage{bm}
\usepackage{color}
\usepackage{hyperref}
\usepackage{caption}
\usepackage{breakurl} 
\hypersetup{breaklinks=true}
\usepackage{multirow}
\usepackage{ctable}
\usepackage[lofdepth,lotdepth]{subfig}

\def\be{\begin{equation}}
\def\ee{\end{equation}}
\def\bea{\begin{eqnarray}}
\def\eea{\end{eqnarray}}

\begin{document}

\title{Electron spin-flip correlations due to nuclear dynamics in driven GaAs double dots}

\author{Arijeet Pal}
\affiliation{Department of Physics, Harvard University, Cambridge, MA, 02138, USA}
\affiliation{Rudolf Peierls Center for Theoretical Physics, University of Oxford, 1 Keble Rd, Oxford OX1 3NP, UK}

\author{John M. Nichol}

\author{Michael D. Shulman}

\author{Shannon P. Harvey}

\affiliation{Department of Physics, Harvard University, Cambridge, MA, 02138, USA}

\author{Vladimir Umansky}

\affiliation{Braun Center for Submicron Research, Department of Condensed Matter Physics,
Weizmann Institute of Science, Rehovot 76100 Israel}

\author{Emmanuel I. Rashba}

\author{Amir Yacoby}

\author{Bertrand I. Halperin}

\email{halperin@physics.harvard.edu}

\affiliation{Department of Physics, Harvard University, Cambridge, MA, 02138, USA}

\pacs{73.21.La, 03.67.Lx}

\begin{abstract}

We present experimental data and associated theory for correlations in a series of  experiments involving repeated Landau-Zener sweeps through the crossing point of a singlet state and a spin aligned triplet state in a GaAs double quantum dot  containing two conduction electrons, which are loaded in the singlet state before each sweep, and the final spin is recorded after each sweep. The  experiments reported here  measure correlations on time scales from 4 $\mu$s to 2 ms. 
When the magnetic field is aligned in a direction such that spin-orbit coupling cannot cause spin flips, the correlation spectrum has  prominent peaks
centered at zero frequency and at the differences of the Larmor frequencies of the nuclei,  on top of a frequency-independent background.   When the spin-orbit field is relevant, there are additional peaks, centered at the frequencies of the individual species.  
A theoretical model which neglects the effects of high-frequency charge noise 
correctly predicts the positions of the observed peaks, and gives a reasonably accurate prediction of the size of the frequency-independent background, but gives  peak areas that are larger than the observed areas by a factor of two or more.  The observed peak widths are roughly consistent with predictions based on nuclear dephasing times of the order of 60 $\mu$s.  However,  there is extra weight at the lowest observed frequencies, which suggests the existence of residual correlations on the scale of 2 ms. We speculate on the source of these discrepancies.

\end{abstract}
\maketitle

\today
 
\section{Introduction}
\label{sec:intro}

Nuclear spins in solid state systems provide a rich platform to study quantum many-body dynamics. The coupling of the electrons to the underlying nuclear environment plays an important role in spintronics \cite{Zutic2004} and utilization of electron spins for quantum computation \cite{Petta2005, Nowack2007, Kloeffel2013}. More generally, the interaction between a driven electron-spin qubit and its \emph{many-body environment} leads to complex dynamical phenomena which are absent if the system is assumed to be at equilibrium with its environment \cite{Chekhovich2013,Petta2008,Foletti2009,BluhmPRL2010,Vink2009,Tartakovskii2007,Latta2009,Lai2006,Ono2004,Greilich2007,Eble2006, Mehl2014, chesi2015, Rancic2014}. Thus, a qubit can be utilized as a probe for studying out-of-equilibrium physics in interacting quantum systems. Furthermore, understanding the rich \emph{system-environment} dynamics is essential for physical implementations of fault-tolerant quantum information processing\cite{Taylor2005}. 

In semiconductor quantum dots the qubit is defined in terms of confined single or multi-electron states in a two-dimensional electron layer confined in a heterostructure.  The singlet ($S$) and $S_z=0$ triplet ($T_0$) states of two electrons in a double quantum dot has proven to be a promising candidate for quantum information processing \cite{Petta2005,Foletti2009,Shulman2012}. The wave functions of the electrons are typically spread over $\sim 100$ nm and the hyperfine interaction between the electrons and their nuclear environment may include several million nuclei. Although the fluctuations in the nuclear spin environment act as a source of decoherence for the $S$-$T_0$ qubit, the difference in polarization of the nuclear spins between the dots of a double quantum dot (DQD) has been usefully exploited to produce rotations around an axis of the Bloch sphere of the qubit. Thus, the stable controllability of the nuclear field gradient is imperative for the efficient control of the qubit. This has been experimentally achieved by protocols to control the state of the nuclear spins through the hyperfine coupling between the electronic and nuclear degrees of freedom \cite{Foletti2009,BluhmPRL2010,Shulman2014}.

Because the energy scale of the interaction between nuclei is much weaker than their hyperfine interaction with the electrons, the consequent separation of time scales allows one to perform high-fidelity quantum control of the $S$-$T_0$ qubit, despite the fluctuating nuclear environment. In this article we shall focus on the anti-crossing between the singlet $S$ and $S_z=+1$ triplet ($T_+$) states of the electrons,  which is utilized for polarizing the nuclear spins. As the gate-voltage is swept through the $S$-$T_+$ anti-crossing, an electron spin can be flipped either by spin-orbit (SO) or hyperfine (HF) interaction, and  in either case, the electron system, starting in the $S$ state,   will emerge in the $T_+$ state. (Note that the $T_+$  state has lower energy than that of  $T_-$, because the $g < 0$ in GaAs.) Transitions caused by the HF interaction will also lead to spin flips in the nuclear system, which can lead to a significant change in the nuclear polarization, if the  sweep protocol is repeated a sufficient number of times \cite{Foletti2009}.    Effects of the nuclear hyperfine field on the electronic spin state have also been employed in experiments on electron dipole spin resonance (EDSR) by various authors \cite{Laird2007,Shafiei2013, Tenberg2015}.

In this article, we discuss an  experiment where the electron sweep protocol is repeated 500 times, and the electronic state, singlet or triplet, is measured and recorded after each sweep.  We then calculate and analyze the power spectrum, which characterizes correlations in the triplet return probabilities for pairs of sweeps that are separated by time intervals $t$ between 4 and 2000 $\mu$s.  Non-trivial correlations are to be expected  in these measurements, because the nuclear configuration will evolve on this time scale, primarily because of Larmor precession in the external magnetic field, which occurs at different frequencies for the three nuclear species involved.  
More generally, the study of these correlations provides crucial insights into the many-body dynamics of the coupled electron-nuclear spin system.  As the cumulative effect of 500 sweeps on the nuclear  polarization is too small to have a significant effect on the triplet return probabilities studied in our experiments,  the experiments may be interpreted as a probe of  intrinsic correlations of the system.

In the article we present our experimental  results and we develop a theoretical model to describe measurements.  The  model is based on the central spin problem \cite{Chen2007, Wang2006, Cywinski2009}, where a two level system is coupled to a large number of spins. The two levels in our case are the $S$ and $T_+$ states of two electrons. The collection of nuclear spins is treated within a semi-classical approximation where the Overhauser fields due to a mesoscopic aggregate of spins can be treated as a set of Gaussian random variables.  In the absence of SO, when the electron spin-flip probability is small, the correlations are dominated by spectral peaks centered at zero frequency and  at the differences of nuclear Larmor frequencies of any two species.
In the presence of SO, there are additional peaks at the individual Larmor frequencies, which are produced due to the interference of the static SO term and the nuclear precession.  The center positions of the correlation peaks extracted from the experiments are in very good agreement with the values predicted from the known Larmor frequencies. The power spectrum also has a frequency-independent background which is well-predicted by our theory. However, our main  focus  will be on the areas and widths of the peaks. 

A large part of this paper will be devoted to theoretical discussions of the predictions for the triplet-return correlation function and its power spectrum that may be deduced from our model. Because of the Gaussian nature of the nuclear spin fluctuation in the model, predictions for the  triplet-return correlation function can be accurately obtained, if  the model parameters are known.  These parameters include the experimental sweep rates and the  strengths of the spin-orbit and the root-mean square hyperfine fields, as well as assumptions about the time scale and form for decay of correlations in the transverse hyperfine fields for each of the three nuclear species.   

Taking values of the key parameters from previous experiments \cite{Nichol2015},  we find qualitative agreement between the predicted peak areas and the experimental results, but there is a systematic discrepancy in which the measured areas are typically smaller, by a factor of two or more than the predictions of the model.  We believe that the most likely cause for this discrepancy is the effect on the triplet return probability due to high-frequency charge noise on the gates or in the quantum dots themselves. It is clear from previous experiments \cite{Nichol2015}, that depending on the sweep rate across the $S$-$T_+$ anti-crossing, the triplet return probability can be significantly affected by such noise.  Although we do not have a detailed knowledge of the size and frequency dependence of the charge noise that may be relevant for the current experiments, and we have  not conducted a quantitative analysis of the possible effects of charge noise on these experiments,  we do include a qualitative discussion, which supports the hypothesis that charge-noise may be the principal source of the remaining discrepancies between the measured peak areas and the theoretical predictions.  

In our analysis of the experimental data, we find that we can obtain a qualitative  understanding of the widths and shape of the peaks in the correlation spectrum by assuming that the decay of correlations has the form of  Gaussian relaxation functions, with correlation times of order 60 $\mu$sec. 
However, the power spectrum has extra weight at the lowest nonzero frequencies studied in these experiments ($\nu \alt 500$Hz),  which is not explained by the model. We discuss in an Appendix the  form of the nuclear spin correlation functions to be expected if decay is primarily the result of inhomogeneous broadening of the nuclear Larmor frequencies. We find that the relaxation functions should be well described by a Gaussian for times that are not too long, but there will be deviations at larger times, which could  possibly lead to anomalies in the triplet return correlation function at very low frequencies.

The general outline of the paper is as follows. In section II we describe in detail the theoretical model used to describe the coupled dynamics of the electron-nuclear system in the double quantum dot. Section III includes the specific multi-sweep protocol which was implemented experimentally and furthermore derives the predictions from the theoretical model for the frequency spectrum of the correlation function in this scenario. In subsection III~A, we present the predictions of our model for the frequency-independent background contribution to the correlation spectrum. In subsection III~B, the consequences of the model for the principal peaks in the spectrum are evaluated within a linear approximation, applicable for fast sweep rates, where the Landau-Zener triplet return probability is small. In subsection III~C, we introduce a nonlinear approximation, valid for smaller sweep rates, which will prove necessary to make sensible predictions for the experiments under consideration. In subsection III~D, we show that an exact solution of our model is possible  for the time-dependence of the triplet-return correlation function, and we explain how these results can be used to obtain precise predictions for the areas of the leading  peaks in the spectral function, if the model input parameters are known.
In section IV we discuss the implications of charge noise in different frequency regimes and comment on their relevance to current experiments. The theoretical predictions of our model, at our various levels of approximation, are compared with each other and with the results of our experiments in Section V, and our conclusions are summarized in Section VI.    As mentioned above, predictions for the form of relaxation for the nuclear spin correlation function due to inhomogeneous broadening are discussed in an Appendix. 


The present paper has some overlap with a recent publication by  Dickel et al. \cite{Dickel2015}. In particular, the results of our full calculation of  the triplet-return correlation function in the time domain, described in Section III D below, coincide with theoretical results described in that publication.  Dickel, et al. also present experimental measurements in the time domain, which agree, at least qualitatively with their theoretical predictions.  By contrast, in the present paper, we present results of an extensive series of measurements and associated theoretical predictions, analyzed in the frequency domain,  which 
enables us to determine separately the effects of hyperfine and spin-orbit coupling on spin-flip correlations in the system.

\section{Multi-sweep experiments and theoretical model}
\label{sec:model}

The system of interest is a double quantum dot containing two electrons. The diameter of the dots in the device used for the experiments is around $\sim 100$ nm. For the temperatures relevant to this work only the lowest lying orbital state of the dots has any significant probability of occupation. Therefore, each dot could either be doubly $\left((0,2)/(2,0)\right)$ or singly $(1,1)$ occupied by the two electrons, although due to Pauli's exclusion principle the spin component of the $(0,2)$ (or $(2,0)$) state is forced to be a singlet. On the other hand, the $(1,1)$ state has no such constraint. The spin component of the electronic wave function is defined in terms of the projection of the spin along the externally applied uniform magnetic field, which we define to be the z-direction. The singlet and triplet states take their canonical form 
\begin{eqnarray}
|S \rangle &=& \frac{1}{\sqrt{2}} \left( |\uparrow_1 \downarrow_2 \rangle - |\downarrow_1 \uparrow_2  \rangle  \right) \\ 
|T_+ \rangle &=& |\uparrow_1 \uparrow_2 \rangle \\
|T_0 \rangle &=& \frac{1}{\sqrt{2}} \left( |\uparrow_1 \downarrow_2 \rangle + |\downarrow_1 \uparrow_2  \rangle  \right) \\
|T_- \rangle &=& |\downarrow_1 \downarrow_2 \rangle 
\end{eqnarray}
in terms of the projections of the spins of individual electrons.

\begin{figure}[!hbtp]
\includegraphics[width=3.0in]{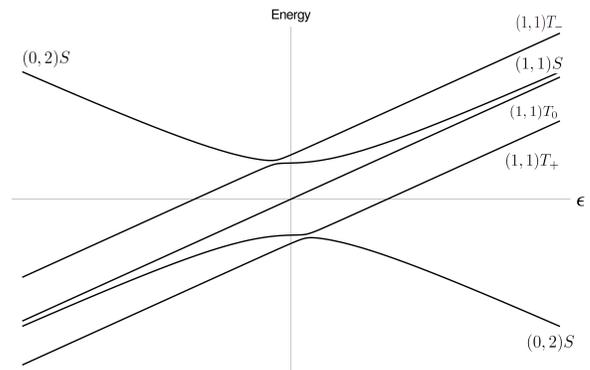}
\caption{Energy level diagram of two electrons in a DQD as a function of detuning $\epsilon$}
\label{fig:energy_diag}
\end{figure}

The energy-level diagram of the system, as a function of voltage difference between the dots (detuning $\epsilon$), is shown in Fig. \ref{fig:energy_diag}. 
In the absence of spin non-conserving terms in the Hamiltonian, the singlet $S$  and triplet $T$ sectors are decoupled from each other. The uniform magnetic field splits the triplet states in energy, producing  gaps equal to the net Zeeman energy of the electrons between   the   $m_s=\pm 1$ triplet states of the electrons $T_\pm$, and the $m_s=0$ triplet $T_0$ state. 
 At any given positive detuning, the electronic states in the singlet sector are an admixture of $(0, 2)$ and $(1,1)$ states due to the tunneling \cite{Fasth2007, BrataasPRB2011, Rudner2010}. On the other hand, due to the total $S_z$ conservation tunneling has no influence on the triplet sector. 

The electron spin non-conserving terms in GaAs semiconductors are due to the nuclear hyperfine  and spin-orbit interactions. They couple the singlet and triplet subspaces of the two electrons. There are several factors contributing to the spin-orbit effect experienced by the electron confined in the dots, like the shape of the dots and the tunneling between them, the orientation of the dots with respect to the crystallographic axes, the spin orbit length of the host material, and magnetic field. The two-dimensional electron gas  is fixed to lie in the $(100)$ plane of the GaAs crystal in the experiments, but  the direction and strength of the in-plane magnetic field are controllable. The magnitude of the spin-orbit interaction strength can be varied by changing the direction of the magnetic field and is given by \cite{Stepanenko2012}
\begin{equation}
\label{vsophi}
v_{so} = |\tilde{v}_{so} \sin \phi | ,
\end{equation}
where $\phi$ is the angle between the in-plane magnetic field and the spin-orbit field direction,  and $\tilde{v}_{so}$ is the strength of the spin-orbit term at the $S-T_+$ crossing point. (In our experiments, the axis of the DQD is either in the $[011]$ or  $[01\bar{1}]$ direction, and the spin-orbit direction is perpendicular to the DQD axis.) The value of  $\tilde{v}_{so}$ will depend on the details of the quantum dot system, and on the magnitude of the applied magnetic field, but not its direction in the plane, as long as the Zeeman energy is much stronger than $\tilde{v}_{so}$. For the relevant size of the gate-defined quantum dots, the electronic wave function is typically spread over $10^6-10^7$ lattice sites. The contact hyperfine interaction with the nuclear spins of $^{69} \text{Ga}$, $^{71}\text{Ga}$, and $^{75}\text{As}$ generates an effective spin-spin interaction between the nuclei and the electrons given by  
\begin{equation}
\hat{H}_{hf}= V_s \sum_{\lambda} A_{\lambda} \sum_{j \in \lambda} \sum_{m=1, 2} \delta \left( \mathbf{R}_{j \lambda}-\mathbf{r}_m \right) ( \mathbf{I}_{j\lambda} \cdot \mathbf{s}_m )
\end{equation}
where $\lambda$ represents the nuclear species, $j$ is the position of the nucleus, and $m$ is the electron index. $V_s$ is the volume per nuclear spin in GaAs. $\mathbf{I}_{j \lambda}$ is the nuclear spin of species $\lambda$ at site $j$ and $\mathbf{s}_m$ is the spin of the $m^{\text{th}}$ electron. The gradient in the transverse component of the nuclear Overhauser field couples the $S$-$T_+$ states and produces an anticrossing  ($\Delta_{ST_+}$) at a particular value of detuning $\epsilon_0$ fixed by the magnetic field. Thus, the direction and magnitude of the external magnetic field serves as a convenient experimental control to tune the $S$-$T_+$ anticrossing between spin-orbit-dominated and hyperfine-dominated regimes. 

It is convenient to define a set of quantities
\begin{eqnarray}
v_n &=& \langle \Psi_S | \hat{H}_{hf} | \Psi_{T_+}  \rangle = \sum_\lambda v_\lambda , \\
v_\lambda &=& V_s  A_{\lambda} \sum_{j \in \lambda} \rho ( \mathbf{R}_{j \lambda}) I_{j \lambda} ^{+} ,
\end{eqnarray}
where $I_{j \lambda} ^{+} \equiv ( I_{j \lambda} ^{x} + i I_{j \lambda} ^{y} )/\sqrt{2}$ and $\rho$ is the hyperfine coupling amplitude defined in terms of the electronic orbital $S$ and $T$ states, {\it viz.} 
\begin{equation}
\rho(\mathbf{R})= \int d^3 \mathbf{r}_2 \psi_T^* (\mathbf{R}, \mathbf{r}_2) \psi_S (\mathbf{R}, \mathbf{r}_2) ,
\label{eq:rho_theta}
\end{equation}
where, $\psi_S$ and $\psi_T$ are the orbital parts of the eigenfunctions $\Psi_S$ and $\Psi_{T+}$.
Although   $I_{j \lambda} ^{+} $ for an individual nuclear spin is an operator that should be treated quantum mechanically, the quantities $v_\lambda$ are each  the sum of very many such variables, and they may be treated, with high accuracy, as classical complex amplitudes, which evolve in time as the nuclei precess about the applied magnetic field.

The effective electronic Hamiltonian at the $S$-$T_+$ anticrossing may then  be written in the form
\begin{equation}
H^{\left( ST_+\right)} = 
\begin{pmatrix}
\epsilon_S & v \\
v^* & \epsilon_{T_+} - \Sigma_{\text{hf}} 
\end{pmatrix} \, ,
\end{equation} 
where  
\be
\label{v=vsovn}
v = v_{so} + v_n,
\ee
and  $\Sigma_{\text{hf}}$ is the sum of the z-components of  
the Overhauser fields on the quantum dots, which gives  rise to a shift in the energy of the triplet state.
Specifically, we may write
\begin{equation}
\Sigma_{\text{hf}} = V_s \sum_{\lambda} A_{\lambda} \sum_{j \in \lambda} \zeta(\mathbf{R}_{j \lambda}) I^z_{j \lambda}
\end{equation}
where the hyperfine amplitude in the triplet state is 
\begin{equation}
\zeta (\mathbf{R}) = \int d^2 \mathbf{r}_2 |\psi_T (\mathbf{R}, \mathbf{r}_2)|^2  .
\label{eq:zeta}
\end{equation}
 Without the loss of generality, the spin-orbit interaction $v_{so}$ will be chosen to be real.

In the experiment the electrons are loaded in the singlet state of the right dot ($(0,2)S$) following which the voltage is swept through the $S$-$T_+$ anticrossing. Assuming that one can neglect effects of high-frequency charge noise, the choice of a linear sweep protocol, $E_S ( t ) - E_{T_+} ( t ) = \beta t/ \hbar$, maps the problem to the famous Landau-Zener case \cite{Landau1977,Zener1932,Stuckelberg1932,Majorana1932} which gives the probability of the transition from the $S$ to $T_+$ state to be 
\begin{equation}
P_{LZ}=1-e^{-2 \pi \gamma}
\end{equation}
for initial ($T_i$) and final ($T_f$) times tending to $-\infty$ and $+\infty$ respectively, and where 
\begin{equation}
\gamma= |v|^2 / \beta.
\end{equation}
In using this relation, for each sweep,  we evaluate $v$ and $\gamma$ at the instant of time when the gate voltage passes through the level-crossing point, where the singlet and $T_+$ states would be degenerate in absence of $v$.     We have assumed that we can neglect the precession of the nuclei during the course of a single sweep, which  should be a good  approximation for the experiments under consideration.   The precession frequencies of the nuclear species are in the range of a few MHz. The duration of the sweep is varied between $50 - 700 \text{ ns}$ to adjust the average $P_{LZ}$.  However, the nuclear configuration is only important during the shorter time interval, when the system is close  enough to the crossing point for an electron spin-flip to occur.  
The total range in the energy difference  $E_S - E_{T^+}$    during a sweep  is $\sim  2 \pi \hbar \times 2 \text{ GHz}$, but the range where spin-flips can occur is  when $|E_S - E_{T^+}| \alt  |v| \approx 2 \pi \hbar \times \left( 10-100 \right) $ MHz.  

In practice, there may be important corrections to the Landau-Zener transition probability due to charge noise in the sample or on the gates.  We shall discuss effects of charge noise in Section V below, but we ignore them for the moment.  

In our experiments, the gate voltage is returned rapidly, to avoid $S$-$T_+$ transitions, to the $(0,2)$ side after each Landau-Zener sweep, the electronic spin state is measured using the spin-blockade technique \cite{Ono2002,Barthel2009}, and 
the outcome is recorded. 
After this measurement,  the electronic state is reinitialized for the next sweep by loading the electrons in the $(0, 2)$ singlet. Successive sweeps are separated in time by a precise time interval $\tau$ = 4$\mu$s, which includes the duration of a sweep as well as the waiting period between sweeps, during which the nuclear spins undergo free Larmor precession. Over the longer time scale of many sweeps, the nuclear spins also exhibit energy and phase relaxation due to nuclear dipole-dipole interactions and other mechanisms,  which we shall take into account in an approximate way.

The measurements were carried out in a series  of ``runs", each consisting of  $N_\tau = 500$  successive sweeps, labeled by $1\leq p \leq N_\tau$,  with sweep times separated by $\tau$ = 4 $\mu$s.   This protocol was repeated $288$ times, with a waiting period of $7.2$ milliseconds between successive runs. At the end of each set of 288 runs, a halt of $90$ seconds is implemented during which all components of the nuclear spins are expected to reach back to equilibrium. This whole procedure was then repeated $50$ times. 
These waiting  periods are sufficiently long that at the beginning of each run, at least the transverse components of the nuclear spin configuration can be assumed to be in a random state sampled from the thermal ensemble,  so  the $288 \times 50=14,400 $ experimental runs may be considered as different realizations of the same ensemble.


For each sweep $p$, we define a variable $\chi_p$ which is equal to 0 or 1 
depending on whether the electron state 
 has flipped from $S$ to $T_+$ or not.  We may then define a spin-flip probability $\langle \chi_p \rangle$ and a correlation function
\begin{equation}
\label{Cchi}
C_\chi(p, q)  = \langle \chi_p \chi_q \rangle, 
\end{equation} 
where $1\leq p,q \leq N_\tau$, and the angular brackets indicate an average over the 14,400 runs.  Analysis of this correlation function will be the main focus of this paper.

The electron spin-flip probability in any given sweep depends on the orientations of the nuclei at the time $t_p$ of that sweep.  As remarked above, the  distribution of the  nuclei before the first sweep in a run should be given by the thermal equilibrium distribution of the nuclei in the applied magnetic field.  For the temperatures and fields relevant to these experiments, the net polarization in the z-direction will be very small compared to  the maximum possible polarization of the nuclei, so that the distribution of  perpendicular spin components should be essentially the same as in an equilibrium ensemble at zero magnetic field.
During the course of 500 sweeps, there may be a change in the z-polarization of the nuclei
due to the effects of dynamic nuclear polarization (DNP), but the polarization will still be very small compared to the maximum polarization.
Therefore, for any single sweep the probability distribution of $v$ should be the same as in thermal equilibrium.

Since the complex variable $v_n$ is the sum  of small contributions from a very large number of nuclei, it is clear that the equilibrium distribution will have the form of a Gaussian, whose form is completely determined by its first and second moments. Since the orientations of different spins are uncorrelated, it is easy to see that $\langle v_n \rangle = 0$, and $\langle |v_n|^2 \rangle = 2 \sigma^2$, 
where $\sigma^2= \sum_\lambda  \sigma_\lambda^2$ and 

\begin{equation}
\sigma_\lambda^2 = n_\lambda A_\lambda^2 \frac{I_\lambda (I_\lambda +1) }{3} 
V_s \int |\rho(\mathbf{R})|^2d^3\mathbf{R} ,
\end{equation}

where $I_ \lambda$ is the spin and  $n_\lambda$ is the fractional abundance of species $\lambda$. For GaAs,  $n_\lambda = $ 0.5, 0.2  and 0.3,  for $^{75}$As,  $ ^{69}$Ga, and $ ^{71}$Ga respectively, while   $I_\lambda = 3/2$ for all species. 
The coupling constants, measured in $\mu$eV  are 
$A_{^{75}As} = 46$, 
$A_{^{69}Ga} = 38.2$, and $A_{^{71}Ga} = 48.5$.  
The quantity 
$N_n \equiv  \left( V_s  \int |\rho(\mathbf{R})|^2d^3\mathbf{R} \right)^{-1} $
may be interpreted as the effective number of  nuclei contributing to the transverse hyperfine field $v_n$.

If we regard the real and imaginary components of $v_n$ as a two dimensional vector $\vec{v}_n$,  the probability distribution of $\vec{v}_n$  may be written as
\begin{equation}
\label{gaussianBH}
p(\vec{v}_n) = \frac {1}{2 \pi \sigma^2}  e^{-|v_n|^2 / 2 \sigma^2} .
\end{equation}
Since the complex  amplitudes $v_\lambda$ are themselves each a sum of contributions from a large number of nuclei, their individual  thermal distributions  are also Gaussians, with $\sigma_\lambda^2$ replacing  $\sigma^2$ in the formula above.

We shall also be interested in the joint  probability distributions of $v_n$ at several different times. 
Under the influence of the applied magnetic field, the macroscopic spins undergo Larmor precession. At the same time the collection of nuclear spins experience energy and phase relaxation due to dipolar and quadrupolar interactions. The time scale for phase diffusion in nuclear spins in GaAs is of the order $100 \mu s$ while that of spin or energy diffusion can be of the order seconds \cite{Reilly2008}.  Since the value of $v_n$ at each time is the sum of contributions from very many nuclei, the joint distribution function of $v_n$ at two different times is again a Gaussian distribution.  Consequently,  the distribution is  completely determined by its second-order correlations.  

\section{Correlations in $S$-$T_+$ sweeps}

The off-diagonal matrix element coupling the $S$ and $T_+$ states has a time-independent part due to the spin-orbit effect and a time-dependent contribution from the transverse components of nuclear spins of the various species, given by   
\begin{eqnarray}
\label{Om}
v_n (t) 
&=& \sum_{\lambda} v_{\lambda}(t) \nonumber \\
&\equiv& \sum_{\lambda} \Omega_{\lambda} (t) e^{-2 \pi i \nu_{\lambda} t}
\end{eqnarray}
where $\nu_{\lambda}$ is the Larmor frequency of species $\lambda$ and the amplitude $\Omega_{\lambda}$ is assumed to vary only slowly, on a time scale of order  100$\mu$s. It is the interference of the terms of different frequencies  contributing to the $S$-$T_+$ matrix element that is the source of the interesting temporal correlations in the electron spin-flip probability $P_{LZ}$.  

Let us write the two-time correlation function for $\Omega_\lambda(t)$ in the form
\begin{equation}
\langle \Omega_{\lambda} (t) \Omega^*_{\lambda'} (t') \rangle = 
2 \delta_{\lambda \lambda'}  \, \sigma_\lambda^2 \,
g_\lambda(t-t') ,
\label{glambda}
\end{equation}
where $g_\lambda(0) = 1$, and $g_\lambda(t) $ decays to zero on a time scale   
 $\tau_{\lambda}$, which  is the relaxation time of species $\lambda$ arising from interactions in the nuclear spin system, etc.  Here, we are assuming that the fluctuations in the nuclear orientations perpendicular to the applied magnetic field can be treated as a stationary stochastic process, which will not be  significantly affected by the Landau-Zener process within a sequence of 500 sweeps.
Motivated by experimental observations, we assume here  a simple  Gaussian form for $g_\lambda$:
 \begin{equation}
g_\lambda(t)   =  e^{- t^2 / 2 \tau_{\lambda}^2} .
\label{eq:GaussDecay} 
\end{equation}
A discussion of  reasons for the (approximate) validity of this assumption, and of possible consequences of deviations from the assumed Gaussian behavior, will be given in  the Appendix.

We now turn to predictions of our model for the correlation function $C_\chi$ defined in (\ref{Cchi}).
Suppose that the nuclear configurations  at the two times $t_p$ and $t_q$ are known, so that the  corresponding LZ probabilities are also known.    Then the conditional expectation value of the product $\chi_p \chi_q$ will be given by 
 \begin{align}
\overline{\chi_p \chi_q} = P_{LZ} (t_p) P_{LZ} (t_q) \left( 1- \delta_{pq}\right) + P_{LZ} (t_p) \delta_{pq} ,
\label{eq:ChiCorr}
\end{align}
since the outcomes $\chi_p$ and $\chi_q$ are stochastic quantities that are independent if and only if $p \neq q$.
If we now average this result over all possible initial conditions of the nuclei, and take into account the effects of random dephasing between the two times $t_p$ and $t_q$, we obtain the result 
\begin {equation}
C_\chi(p,q) = f_B  \,\delta_{pq} + \langle  P_{LZ} (t_p) P_{LZ} (t_q) \rangle ,
\end{equation}
where
\begin{equation}
\label{fb1}
f_B =  \langle \,  P_{LZ} (t_p)  - [ P_{LZ} (t_p)] ^2  \, \rangle .
\end{equation}
As argued above, this expectation value should be essentially independent of $p$. We remark that the term proportional to $f_B$ is a quantum stochastic effect, which reflects the random outcome for the value of $\chi_p$, even when the probability $P_{LZ}$ is specified. This will lead to a frequency-independent background contribution to the Fourier transform of $C_\chi$.

In practice, it will be  most convenient to  work  with a Fourier expansion of  $\chi_p$ and   to discuss the power spectrum of $C_\chi$.    We define
\begin{equation}
\tilde{\chi}_{n} = \sum_{p=-N_{\tau}/2}^{N_{\tau}/2} e^{2\pi i n p} \, \chi_p,
\end{equation}
 where $n$ is an integer, and we  impose the restriction  $-1/2\tau < \nu_n \le 1/2\tau$,  where  $\nu_n$ is the frequency defined by   
\begin{equation}
\nu_n \equiv \frac{n}{N_{\tau} \tau}.
\end{equation}
The power spectrum is then defined  as
\begin{equation}
\label{Fnun1}
F(\nu_n) = \langle {|\tilde{\chi}_{n}|^2}\rangle = \sum_{p,q} e^{2 \pi i n (p-q)} \,C_\chi(p,q) .
\end{equation}

We define a correlation function 
\begin{equation}
\label{f27}
f(t_p, t_q) \equiv  \langle  P_{LZ} (t_p) P_{LZ} (t_q) \rangle = f(t_p-t_q),
\end{equation}
which depends only on the time separation $(t_p-t_q)$, and should be a continuous function of that variable. (This is because the values of $v_\lambda$ evolve continuously in time, and are not affected by any intervening Landau-Zener sweeps on the time scale we are considering.)  For times large compared to the dephasing times $\tau_\lambda$, the function $f(t)$ will approach a limit,
\begin{equation}
\label{ftoinfty}
f(t) \xrightarrow{|t| \rightarrow \infty} f_{\infty} = \langle P_{LZ} \rangle^2.
\end{equation}
Taking the Fourier transform of $f(t)$, after subtracting the infinite time limit, we define a function
\begin{equation}
S(\nu) \equiv \int^{+\infty}_{-\infty} dt \, e^{-2 \pi i \nu t} \left( f(t) - f_{\infty} \right) \, .
\end{equation}

We now wish to relate the experimentally observed power spectrum  $F(\nu_n)$ to the function $S(\nu)$.  The functions differ for three reasons: because $F$ includes a contribution from the background term $f_B$ which is omitted from $S$,  because the experimental measurements  are restricted to a discrete set of time steps rather than as a continuous function of time, and because the measurements are restricted to a finite time interval $N_\tau \tau$.  This last restriction should be unimportant, provided that the time interval $N_\tau \tau$ is large compared to all of the correlation times $\tau_\lambda$. The contribution of $f_B$  can be added explicitly, and the difference between the discrete sum and the continuous integral can be handled by use of the Poisson sum formula.  The result is 
\begin{equation}
\label{Fnun2}
F(\nu_n) = N_\tau^2 f_\infty \delta_{n 0}  + N_\tau f_B + \tilde{F}(\nu_n),
\end{equation}
\begin{equation}
\label{Fnun3}
\tilde{F}(\nu) = \frac{N_\tau}{ \tau} \sum_{l= - \infty}^{\infty} S\left( \nu + \frac{l}{\tau} \right) .
\end{equation}

\subsection{Background $f_B$}

The frequency-independent background of the power spectrum,  $N_\tau f_B$, may be computed by performing the  average indicated in       
Eq. (\ref {fb1}) over the nuclear distributions given by Eq.~(\ref{gaussianBH}): 
\begin{eqnarray}
f_B &=&  \int d ^2 \vec{v}_n \,  p(\vec{v}_n ) \, P_{LZ} \left( 1 - P_{LZ} \right) \nonumber \\
     &=& \langle e^{-2 \pi \gamma} \rangle - \langle e^{-4 \pi \gamma} \rangle .
\label{eq:S_woso_B}
\end{eqnarray}
In the absence of  spin-orbit interactions, the integrals are simple Gaussian integrals, and one obtains, after a small amount of algebra:
\begin{equation}
\label{fb2}
f_B = \frac{\langle P_{LZ} \rangle \left( 1 - \langle P_{LZ} \rangle \right)}{1+\langle P_{LZ} \rangle} \, \,,
\end{equation}
where 
\begin{equation}
\label{plz34}
\langle P_{LZ} \rangle =
\frac{\frac{4 \pi}{\beta} \sigma^2}{1+\frac{4 \pi}{\beta} \sigma^2}  
\end{equation}
and $\sigma^2 \equiv \sum_{\lambda} \sigma_{\lambda}^2$.

The background in the presence of spin-orbit interaction can again be calculated at all orders in $\gamma$.  The first  average in Eq.(\ref{eq:S_woso_B}) is now given by  
\begin{align}
\langle e^{-2 \pi \gamma} \rangle &= \prod_{\lambda} \int d^2 \vec{v}_{\lambda}  \, p \left(  \vec{v}_{\lambda}    \right) \nonumber \\
& \times \exp \left(-\frac{2\pi}{\beta} \left( \left( v_{so} + \sum_{\lambda} v_{\lambda, r}\right)^2 + \left( \sum_{\lambda} v_{\lambda, i}\right)^2  \right) \right) \nonumber\\
&= \frac{1}{1 + \frac{4 \pi \sigma^2}{\beta}} \exp \left( - \frac{\frac{2\pi v^2_{so}}{\beta}}{1+\frac{4\pi \sigma^2}{\beta}} \right) 
\label{eq:S_wso_B2}
\end{align}
where $v_{\lambda, r/i}$ represents the real and imaginary part of  $v_\lambda$.  A similar calculation gives
\begin{equation}
\langle e^{-4 \pi \gamma} \rangle = \frac{1}{1 + \frac{8 \pi \sigma^2}{\beta}} \exp \left( - \frac{\frac{4\pi v^2_{so}}{\beta}}{1+\frac{8\pi \sigma^2}{\beta}} \right)
\end{equation}

\subsection{Linear Approximation for $S(\nu)$}
\label{subsec:linear}

We now discuss predictions for the function $S(\nu)$ by first considering some  simple cases.   We begin by considering a linear approximation, which is valid in the regime of fast Landau-Zener sweeps, where $2 \pi \gamma \ll1$. 
In this  regime the Landau-Zener probability is small, and  it can be expanded in a power series in $2 \pi \gamma:$
\begin{eqnarray} 
P_{LZ} (t_p) &=& 1-e^{-2 \pi \gamma} \nonumber \\
&\approx& 2 \pi \gamma + \ldots = \frac{2 \pi}{\beta} |v (t_p)|^2  + \dots
\label{eq:PLZ_won}
\end{eqnarray}
where $t_p=p \tau$ is the time of the $p$-th sweep.   

\subsubsection{\textbf{Case $v_{so} = 0$}}
In the absence of SOI,  only the nuclear spin terms are responsible for correlations in the electron spin-flip probability. For small $P_{LZ}$, the Fourier transform of the lowest order term in $P_{LZ}$ is given, for $\nu \neq 0$,  by
\begin{align}
&\frac{S (\nu)}{\left( 2 \pi/\beta \right)^2 } = 
\int_{-\infty}^{+\infty} ds e^{-2 \pi i \nu s} [f(s)- f_\infty] \nonumber \\
& \approx \sum_{\lambda \lambda' \mu \mu'} \int ds e^{-2 \pi i \nu s}  e^{-2 \pi i (\nu_{\lambda}-\nu_{\lambda'})s} e^{+i (\nu_{\mu}-\nu_{\mu'})s} \nonumber \\
& \times \langle  \Omega_{\lambda} (t) \Omega^*_{\lambda'} (t)  \Omega_{\mu} (t+s) \Omega^*_{\mu'} (t+s)\rangle \,.
\label{eq:S_woso}
\end{align}
On averaging over the nuclear spin configuration, the power spectrum has peaks at frequencies equal to the differences of the Larmor frequencies of any two of the species. In the absence of nuclear spin relaxation, these peaks are delta functions in frequency, but as we include nuclear relaxation phenomenologically, these peaks broaden and develop a finite line-width consistent with a Gaussian decay of correlations given by Eq. (\ref{eq:GaussDecay}). 
On taking into account the Gaussian decay in time, the resulting expression is 
\begin{equation}
 S(\nu)  =  \left( 4 \pi/\beta \right)^2  \sum_{\lambda , \mu} \sigma_{\lambda}^2 \sigma_{\mu}^2  G_{\lambda \mu} (\nu)  ,
\label{eq:Phi}
\end{equation}
where $G_{\lambda \mu}(\nu)$ is a Gaussian of unit area, given by 
\begin{equation}
G_{\lambda \mu} (\nu) = \sqrt{\pi} \tau_{\lambda \mu} \, \exp \left(- \pi^2 \left( \nu + \nu_{\lambda} - \nu_{\mu} \right)^2 \tau^2_{\lambda \mu} \right) \label{eq:G_la} ,
\end{equation}
\begin{equation}
\label{tlmtltm}
1/\tau^2_{\lambda \mu} \equiv (1/\tau^2_{\lambda} + 1/\tau^2_{\mu})/2.
\end{equation}  
 The Gaussian peak around zero frequency  receives contributions from all the three species additively, and thus is much stronger than the peaks at difference frequencies.
If we substitute the expression (\ref{eq:Phi}) into (\ref{Fnun2}), we obtain an approximate expression for the power spectrum, which we can compare with experiments.  

In Figure 2, we show experimental data (black curve) for $F (\nu_n)$, with  data for the singular point  $\nu_n=0$ omitted, taken at two values of the applied magnetic field $B$. The magenta curve is an empirical fit of the data to a set of Gaussian peaks, sitting on top of a frequency-independent background.
Data are shown only  in the positive half Brillouin zone,  $0<\nu_n \leq (2 \tau) ^{-1}$ = 125 kHz, as  the spectrum depends only on $|\nu_n|$. 

In each plot, one sees clearly three Gaussian peaks  centered at non-zero frequencies, as well as the positive half of a quasi-Gaussian peak centered at $\nu=0$, all of which sit on top of a frequency-independent background.   The vertical lines are drawn at the three difference frequencies $(\nu_\lambda - \nu_\mu)$ mod $(1/\tau)$ which fall in the positive half Brillouin zone.  It can be seen that the centers of the Gaussian peaks agree with the positions of the vertical  lines  to a high degree of accuracy.  We defer, until Sec. V, a more detailed comparison between theory and experiment, including the areas under the peaks, the relative widths of the peaks, and the height of the background.

In addition to  the  quasi-Gaussian peak around $\nu=0$, the data shows enhanced values of the spectrum for the lowest non-zero values of  the discrete frequency, particularly at   $\nu_{n=1} = 0.5$ kHz.  This will be discussed further in  Sec.  V.

If one extends the theoretical analysis beyond the first term in the expansion of $P_{LZ }= 1-e^{-2 \pi \gamma}$, one expects to find additional peaks at arbitrary linear combinations of the difference frequencies $\nu_\lambda - \nu_\mu$, reduced to the first Brillouin zone.  However the areas of the higher order peaks will be relatively small for the values of $\gamma$ of interest to us, and the widths of the peaks become larger with increasing order.  It is therefore not surprising that we do not see signs of  higher order peaks in the experimental data.

\begin{figure}[!hbtp]
\subfloat[]{
\includegraphics[width=3.0in]{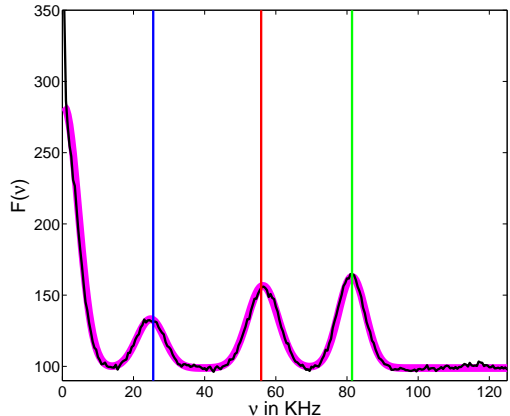}
\label{fig:PLZ_corr_B_0.19_PLZ_0.4}
}
\\
\subfloat[]{
\includegraphics[width=3.0in]{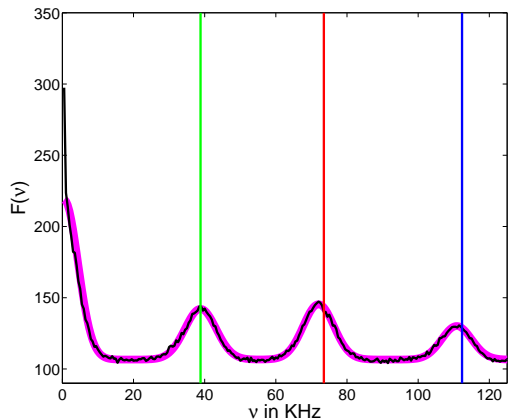}
\label{fig:PLZ_corr_B_0.4_PLZ_0.4}
}
\caption{Figs. \ref{fig:PLZ_corr_B_0.19_PLZ_0.4} and \ref{fig:PLZ_corr_B_0.4_PLZ_0.4}  are the plots of the power spectrum $S$   of the Landau-Zener probability ($\langle P_{LZ} \rangle=0.4$ in both cases), at $B=0.19 \text{ and } 0.4 \text{ T}$, respectively. The black curves are the experimental data while the magenta curves are an empirical fit to a sum of Gaussian peaks sitting on a frequency independent background. Vertical lines are at the difference frequencies between two different species: $\nu_{^{69}Ga}-\nu_{^{71}Ga}$ (blue), $\nu_{^{71}Ga}-\nu_{^{75}As}$ (green), and $\nu_{^{75}As}-\nu_{^{69}Ga}$ (red).}
\label{fig:MultiSweep_PLZ_wo_SOI}
\end{figure}

\subsubsection{\textbf{Case $v_{so} \neq 0$}}

The presence of spin-orbit coupling allows for another mechanism for electron spin-flips besides nuclear spins. In the $S$-$T_+$ matrix element,  the effective SO interaction $v_{so}$,  which depends on the  angle $\phi$ between the spin-orbit field and applied magnetic field according to Eq.~(\ref{vsophi}), can be varied by changing the direction of the field in the plane of the sample. In this regime, the correlations in $P_{LZ}$ receive contributions from the spin-orbit term in combination with the dynamics of the nuclear spins. In the approximation where we keep only the lowest order term in the expansion of 
$(1-e^{-2 \pi \gamma})$, interference of the two effects generates terms proportional to $v^2_{so}$ in the correlation function $S(\nu)$  with peaks at the Larmor frequencies of the individual species, in addition to the terms in Eq.~(\ref{eq:S_woso}). On using the form of the $S$-$T_+$ matrix element, given by
\begin{equation}  
v = v_{so} + \sum_{\lambda} \Omega_{\lambda} e^{-2 \pi i \nu_{\lambda} t} ,
\end{equation}
[{\it cf.} (\ref{v=vsovn}) and (\ref{Om})],     the power spectrum in the presence of SOI acquires an additional term,  so we now have
\begin{equation}
S^{so} (\nu) = S^0 (\nu) + (8 \pi^2/ \beta^2 ) \,  v_{so}^2 \sum_\lambda \sigma_\lambda^2  G_{\lambda} (\nu)  
\label{eq:SpectralFn_PLZ_with_SOI}
\end{equation}
where $S^0$ is the predicted spectrum for $v_{so}=0$, given by Eq. (\ref{eq:Phi}), and 
\begin{align}
& G_{\lambda} (\nu) =
 \int^{+\infty}_{-\infty} ds e^{-2 \pi i \nu s -\frac{s^2}{2 \tau^2_{\lambda}}} \left(  e^{2 \pi i \nu_{\lambda} s} + e^{-2 \pi i \nu_{\lambda} s} \right)\nonumber \\
&=  \sqrt{2\pi}\, \tau_{\lambda}\left(e^{-2 \pi^2 \left(\nu -\nu_{\lambda}\right)^2 \tau^2_{\lambda}} + e^{-2 \pi^2 \left(\nu +\nu_{\lambda}\right)^2 \tau^2_{\lambda}}\right).
\end{align}

Thus, the power spectrum in the presence of SOI, $S^{so}$, has additional  peaks at the bare Larmor frequencies of the three different species given by the functions $G_{\lambda}$. (Of course, in $\tilde{F}$, the bare frequencies $\nu_{\lambda}$ are measured modulo $1/\tau$.) It is interesting to note that the widths of the additional peaks due to the presence of SOI are predicted to be narrower than the peaks at the differences of the Larmor frequencies.

If $v_{so}$ is turned on while the sweep rate is fixed, so that the values of $\sigma^2_\lambda$
are unchanged, the value of $\langle P_{LZ} \rangle$ will increase, as follows from (\ref{v=vsovn}) and (\ref{eq:PLZ_won}).  This will lead to an increase in the weight $N_\tau^2 f_\infty$ of the delta function at zero frequency, which is proportional to $\langle P_{LZ} \rangle^2$, according to (\ref{ftoinfty}). However, the change in $\langle P_{LZ} \rangle$ may be removed by an increase in the sweep rate, if desired.

\begin{figure}[!hbtp]

\subfloat[]{
\includegraphics[width=3.0in]{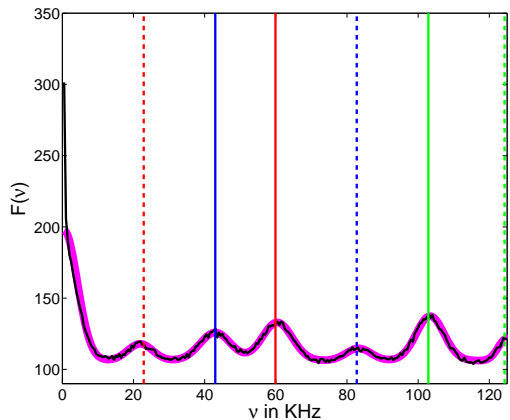}
\label{fig:PLZ_corr_B_0.1_PLZ_0.6_phi5}
}
\\
\subfloat[]{
\includegraphics[width=3.0in]{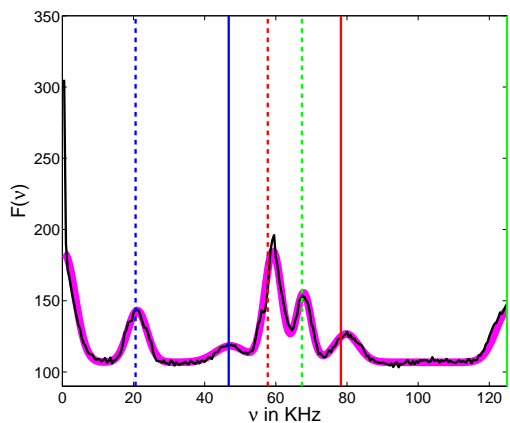}
\label{fig:PLZ_corr_B_0.19_PLZ_0.42_phi10}
}
\caption{(a) Power spectrum of $P_{LZ}$ for $B=0.1 \text{ T}$, $\langle P_{LZ}\rangle=0.6$, and the direction of the field $\phi = 5^{\circ}$. (b)  Power spectrum of $P_{LZ}$ for $B=0.19 \text{ T}$, $\langle P_{LZ}\rangle=0.42$, and the direction of the field $\phi = 10^{\circ}$. The black curves are the experimental data while the magenta curves are an empirical fit to a sum of Gaussian peaks on a frequency-independent background. 
In both the figures solid vertical lines are at the difference frequencies between two different species: $\nu_{^{69}Ga}-\nu_{^{71}Ga}$ (blue), $\nu_{^{71}Ga}-\nu_{^{75}As}$ (green), and $\nu_{^{75}As}-\nu_{^{69}Ga}$ (red), while the dashed vertical lines are at the bare frequencies: $\nu_{^{69}Ga}$ (blue), $\nu_{^{71}Ga}$ (green), and  $\nu_{^{75}As}$ (red).  }
\label{fig:MultiSweep_PLZ_with_SOI}
\end{figure}

Fig.~\ref{fig:MultiSweep_PLZ_with_SOI} presents experimental results for the spectral function $F(\nu_n)$ for two different values of the angle $\phi$, which give rise to increasing values of $v_{so}$. Vertical lines show the positions expected for the bare Larmor frequencies and the difference frequencies, which align extremely well with the positions of the experimental peaks, as expected from our model.   Comparison between predicted and observed peak heights and areas, as well as the frequency independent background, will be discussed in Section V.

\subsection{Nonlinear Approximation for the Peak Areas}
\label{subsec:non-linear}

The formulas for the areas of the Gaussian  peaks, derived in the the previous two subsections,   are correct to lowest order in $\gamma$, i.e., order $\gamma^2$, when $2 \pi\gamma $ is small and $P_{LZ}(\gamma)$ may be adequately  approximated by  $2 \pi \gamma$. This assumption  is  correct when  the sweep rate $\beta$ is sufficiently large.  For the experimental data to be discussed below, however, this linear approximation is not adequate.  

As will be discussed in Subsection III D,  an exact analytic  calculation of the correlation function   $f(t-t') = \langle P_{LZ}(t' ) P_{LZ}(t)  \rangle$ of Eq. (\ref{f27}), correct for arbitrary $\gamma$,     is   possible for our model, in the absence of charge noise, assuming that  the input parameters are known.   However, to extract the areas of the peaks in the frequency domain, it is necessary to take the Fourier transform numerically, and the results are not transparent.
We shall therefore begin by presenting an approximate nonlinear  calculation, which yields transparent analytic results that are a major improvement over the lowest order results, and which also give some physical insight into the size of the necessary corrections.  

We shall be interested here in the areas under the peaks in the correlation function $S(\nu)$,  and we will not pay attention to the detailed line shape.  Our discussions, therefore, will be independent of the precise time dependence of  the correlation functions $g_\lambda(t)$ defined in (\ref{glambda}).

Let us write
\be
\gamma(t) = \tilde{\gamma}(t) + \delta \gamma (t),
\ee
where
\be
\tilde{\gamma} \equiv \beta^{-1} \left( v_{so}^2 + \sum_{\lambda} |v_{\lambda}|^2 \right) ,
\ee
\be
\delta \gamma = \beta^{-1} \left[
v_{so}  \sum_\lambda   ( v_\lambda + v_\lambda^*) +  \sum_{\lambda \neq \lambda '} v_\lambda^*  v_{\lambda '} \right] .
\ee
Then we may expand $P_{LZ}$ as 
\be
\label{Pexp}
P_{LZ}(\gamma) = P_{LZ}(\tilde{\gamma}) + \delta \gamma P'_{LZ}(\tilde{\gamma}) 
+ \frac{1}{2} (\delta \gamma)^2 P''_{LZ}(\tilde{\gamma}) ,
\ee
where our approximation shall consist in omitting terms that are higher order in $\delta \gamma$. 

As in the previous subsections, we assume that $v_\lambda(t)$ is given by (\ref{Om}), where $\Omega_\lambda(t)$ varies slowly in time, with a correlation function of form (\ref{eq:GaussDecay}). The precise form of the  correlation function is not important for the present purposes; we need only assume that  ({\it i}) the $\Omega_\lambda$ are complex variables with  a Gaussian joint probability distribution, ({\it ii}) that there are no correlations  between different species,  and that ({\it iii}) there exists a coherence time $\tau_{coh}$ such that  the correlation function vanishes for $t \gg \tau_{coh}$, but is essentially independent of time for $t \ll \tau_{coh}$.

For $t \ll \tau_{coh}$, we may expand  the correlation function $f(t)$  as 
\be
\label{fexp}
f(t) = f_0 + 2 \sum_\lambda  f_\lambda \cos(2 \pi \nu_\lambda t ) + 
\sum_{\lambda \neq \lambda'}  f_{\lambda  \lambda ' } e^{2 \pi i (\nu_\lambda - \nu_ {\lambda '})t } , 
\ee
where we have omitted terms  containing other combinations of frequencies, which will turn out to be higher order in our expansion.   When we take the Fourier transform of $f$, we find that the terms included in (\ref{fexp})  give rise to narrow peaks in $ S(\nu)$,  centered at frequencies $\pm \nu_\lambda$ or $(\nu_\lambda - \nu _{\lambda'})$, whose areas are given by the coefficients $f_\lambda$ or $f_{\lambda \lambda '}$, respectively.  The width of the peaks are of order  $1 / \tau_{coh}$, but the areas do not depend on 
$\tau_{coh}$.  Similarly, there will be a peak  in $S$  centered at $\nu=0$, with width of order  $1 / \tau_{coh}$, whose area will be equal to $f_0 -f_\infty$.

Now, using the approximation  (\ref{Pexp} ), we find that 
\be
f_0 \approx  \langle  P_{LZ}^2 (\tilde {\gamma}) \rangle +
\langle  P_{LZ} (\tilde {\gamma})   P_{LZ}'' (\tilde {\gamma})  (\delta \gamma)^2 \rangle ,
\ee
\be 
\label{flala}
f_{\lambda \lambda'}  \approx  \beta^{-2}   \langle [ P_{LZ} '  (\tilde {\gamma}) ]^2   |\Omega_\lambda |^2     
|\Omega_{\lambda ' } |^2  \rangle ,
\ee
\be
f_\lambda  \approx  \beta^{-2} v_{SO}^2   \langle [ P_{LZ} ' ( \tilde {\gamma}) ]^2   |\Omega_\lambda |^2     
 \rangle  .
\ee

The above expressions  can be evaluated  using the equalities
\be
\frac {- P_{LZ}'' (\tilde{\gamma})} {4 \pi^2}  = \frac{P_{LZ}' (\tilde{\gamma})} {2 \pi}  = [1- P_{LZ} (\tilde{\gamma})]
= e^{-2 \pi \tilde{\gamma}}.
\ee
 
 It is convenient to define the quantities
 \be
x \equiv  \frac {2 \pi}{\beta}  v_{so}^2 , \, \,\,\,\, y_\lambda \equiv \frac{4 \pi}{\beta} \sigma_\lambda^2 , \,\,\,\,\, 
y \equiv \sum_{\lambda} y_{\lambda} ,
\ee
\be
p_n = \prod_\lambda (1 + n y_\lambda)
\ee
 Then the  results, which one finds after some algebra, ({\it cf.} the calculations in Subsection III A, above), are
 \be 
 f_{\lambda \lambda'}  \approx  \frac{  e^{-2x} y_\lambda  y_{\lambda'}  } 
 {  p_2    ( 1+2 y_{\lambda})(1+2 y_{\lambda'} )      } ,
\ee
\be
f_\lambda \approx  \frac{  e^{-2x} x  y_\lambda } 
 { p_2       (1+2 y_{\lambda})   } ,
\ee
while the two terms contributing to $f_0$ are given by
\be
 \langle P_{LZ}^2 (\tilde {\gamma}) \rangle = 1 - \frac{2 e^{-x}}{p_1} + \frac {e^{-2x}}{p_2}
\ee
\begin{eqnarray}
&&  \langle  P_{LZ} (\tilde {\gamma})   P_{LZ}'' (\tilde {\gamma})  (\delta \gamma)^2 \rangle  \nonumber \\  &=&
\sum_\lambda  \frac { 2  \,x \, y_ \lambda \, e^{-2x}} { p_2   (1+ 2 y_\lambda) } -\sum_\lambda  \frac { 2  \,x \, y_ \lambda \, e^{-x}} {p_1  (1+ y_\lambda) }  
 \nonumber \\  &+&
\sum ^\prime  \frac {    e^{- 2x} y_\lambda y_{\lambda '}} {(1 + 2y_{\lambda''}  )(1 + 2y_{\lambda'}  )^2 (1 + 2y_{\lambda})^2  } 
  \nonumber \\  &-&    \sum^\prime  \frac { e^{-x}   y_\lambda y_{\lambda '}} {(1 + y_{\lambda''} ) (1 + y_{\lambda'}  )^2 (1 + y_{\lambda})^2  } .
\end{eqnarray}
The primes over the last two summation signs signify that the sums are to be taken over $\lambda$ and $\lambda'$, with $\lambda'  \neq \lambda$,  while $\lambda ''$ denotes the third species, not equal to $\lambda$ or $\lambda '$.    

It should be emphasized that the expansion coefficients  $f_0 , f_\lambda, f_{\lambda \lambda'}$, etc., are all independent of the values of the frequencies $\nu_\lambda$  and are well defined, as long as the frequencies are incommensurate  with each other.

As one test of the validity of these  approximations, we may calculate the value of $\langle P_{LZ} \rangle$ using the expansion (\ref{Pexp}): 
\begin{eqnarray}
\label{papprox}
\langle P_{LZ}  \rangle &\approx& \langle P_{LZ} (\tilde{\gamma} ) \rangle + 
\frac{1}{2} \langle P_{LZ}'' (\tilde{\gamma} )  (\delta \gamma)^2   \rangle \nonumber \\ &=&
1 - \frac {e^{-x} }{p_1} -\frac{x e^{-x}}{p_1} \sum_{\lambda} \frac{y_{\lambda}}{1+y_{\lambda}} \nonumber \\ &-& 
\frac{e^{-x}} {2 p_1^2} \sum^\prime   y_\lambda y_{\lambda '} (1+ y_{\lambda ''} )    \,  ,  
\end{eqnarray}
and we may compare the result with the exact answer. As an example, if  we set $v_{so} = 0$, and $y_\lambda = 2/9$ for all three species, the exact value of 
 $\langle P_{LZ} \rangle$ is given by $y /(1+y) = 2/5$, while the number predicted by  Eq. (\ref{papprox}) is
0.3980.    The value of $ \langle P_{LZ} (\tilde{\gamma} ) \rangle$ in this case is 0.4523.

More generally, we expect that the expansion (\ref{Pexp}) should be reasonable as long as the individual $y_\lambda$ are small, even if the sum $y+x$ is not.  

\subsection{Full calculation}

It is convenient to write
\be
f(t) = 2 \langle P_{LZ}\rangle - 1  + P_2(t) 
\ee
where $\langle P_{LZ} \rangle$ is given by Eq.~(\ref{plz34}) and
\be
\label{P2def}
P_2(t) = \langle e^{- 2 \pi \gamma (t)} e^{- 2 \pi \gamma (0)} \rangle
\ee
Since the variables $\Omega_\lambda$ have a Gaussian distribution, this last  expectation value can be expressed as a multivariable Gaussian integral, which can be evaluated by standard methods.

In the regime where $t \ll \tau_{coh}$ , the values of $\Omega_\lambda$  may be assumed to be independent of time, so the evaluations require only integration over three independent complex variables.  Then, in the case where $v_{so} = 0$, the results simplify further to give
\be
\label{P2det}
P_2 = \frac{1}{\det [ M]}
\ee 
 where $M$ is the $3  \times 3$ matrix
 \be
 M_{\lambda \lambda'} (t) = \delta_{\lambda \lambda'} +
 \frac {4 \pi}{\beta} \sigma_{\lambda} \sigma_{\lambda'} \left( 1 + e^{-2 \pi i \left( \nu_\lambda - \nu_{\lambda'} \right) t}\right) ,
 \ee

In the case where $v_{so} \neq 0$,   the result  for  $t \ll \tau_{coh}$  becomes
\be
P_2 = \frac {\exp \, \left(- 2x + 4 \pi  x  \beta^{-1}  \sum_{\lambda {\lambda'} }  \tilde{\sigma} ^*_ \lambda  (M)^{-1} _{\lambda  \lambda'} 
\tilde{\sigma}_{\lambda'} \right) } { \det [M] } ,
\ee
where 
\be
\tilde{\sigma}_\lambda \equiv \sigma_\lambda (1 + e^{-2 \pi i \nu_\lambda t}) .
\ee

The  equations above may be simplified further  by using the results
\begin{widetext}
\begin{equation}
\det[M]=1+\frac{8\pi}{\beta}\sum_\lambda\sigma_\lambda^2+\frac{32\pi^2}{\beta^2}\sum_{\lambda\neq\mu}\sigma_\lambda^2\sigma_\mu^2\sin^2[\pi(\nu_\lambda-\nu_\mu)t] , 
\label{eqR11}
\end{equation}
\begin{equation}
\sum_{\lambda\mu}{\tilde\sigma}_\lambda^*(M^{-1})_{\lambda\mu}{\tilde\sigma}_\mu=4\frac{\sum_\lambda\sigma_\lambda^2\cos^2(\pi\nu_\lambda
t)+\frac{4\pi}{\beta}\sum_{\lambda\neq\mu}\sigma_\lambda^2\sigma_\mu^2\sin^2[\pi(\nu_\lambda-\nu_\mu)t]}{\det[M]}  .
\label{eqR10}
\end{equation}
\end{widetext}

As may be seen from the above equations, in the limit $\tau_{coh} = \infty$, the function $f(t)$ is a quasiperiodic function, with three fundamental frequencies $\nu_j, \, (j = 1,2,3)$, corresponding to the three different values of $\nu_\lambda $.  Then we can expand $f(t)$ in the form
\be
f(t) = \sum f_{m_1 m_2 m_3} e^{- 2 \pi i t (m_1 \nu_1 + m_2 \nu_2 + m_3 \nu_3)} ,
\ee 
where $m_j$ are integers running from $- \infty$ to $\infty$.
The expansion coefficients $ f_{m_1 m_2 m_3}$ may then be obtained by taking the limit $T \to \infty$ of the integral
\be
\frac {1}{2T}  \int_{- T}  ^T  dt f(t) e^{ 2 \pi i t (m_1 \nu_1 + m_2 \nu_2 + m_3 \nu_3)} .
\ee
(In practice, convergence can be improved by using a soft cutoff in the above integration.) The quantities $f_0, f_\lambda$ and $f_{\lambda \lambda'}$ of Eq.~(\ref{fexp}),
which give the areas of the lowest order peaks in $S(\nu)$, are given by the coefficients $ f_{m_1 m_2 m_3}$ with $m_j = 0$ or $\pm1$,  and $|\sum m_j| \leq 1$.

If one wishes to calculate  $P_2(t) $  in the regime of intermediate times, where $t$ is comparable to $t_{coh}$, then the values of $\Omega_\lambda$ at $t$ and $t=0$ should be treated as separate, but correlated, Gaussian variables.  The expectation value in Eq. (\ref{P2def}) would then  be expressed as an integral over a Gaussian function of six complex variables, or twelve real variables. As a simpler alternative, however, 
 one may consider the correlation function $g_\lambda(t)$ as arising from 
an inhomogeneously broadened line, so that  
\be
\sigma^2_\lambda  \,g_\lambda(t) =  \sum_\alpha \sigma^2_{\lambda \alpha} \exp (- 2 \pi i t \delta \nu_{\lambda \alpha}) ,
\ee
where the set of $\delta \nu_{\lambda \alpha}$ denote frequency shifts from the line center, and $\sigma^2_{\lambda \alpha}$ are the corresponding weights.  Then $P_2(t)$ may be evaluated by treating each $\nu_{\lambda \alpha}$ as  arising from a different nuclear species and replacing the indices $\lambda$ and $\mu$
in formulas (\ref{P2det}) to (\ref{eqR10}) by $(\lambda , \alpha)$ and $(\mu, \beta)$.  


 In the regime where $t$ is comparable to $t_{coh}$, the function $f(t)$ is no longer quasiperiodic, so the Fourier transform will  no longer be a sum of sharp $\delta$-functions.

\section{Effects of charge noise}

A 2DEG buried around $100$ nm below the surface of a semiconductor heterostructure is susceptible to charge noise from various possible sources, including  the random two-level systems in adjoining material or through the metal gates on the surface \cite{Dial2013}. Effects of charge noise may be modeled by including fluctuations $\delta \epsilon (t)$  in the detuning parameter  $\epsilon$ relative to the nominal $\epsilon_0 (t)$ specified by the LZ  sweep protocol.
The consequences of these fluctuations will depend on their characteristic frequency.

\subsection{High-Frequency Noise}

Noise fluctuations may  be considered to be ``high frequency'' if they occur at frequencies that are comparable to or larger than the typical value of the
$S$-$T_+$ spitting, $|v|$. 
Effects of high frequency noise were discussed theoretically by 
Kayanuma \cite{Kayanuma1984}, and were discussed more recently in the supplementary material to 
Ref.~\cite{Nichol2015} 
in the context of the present experimental system.  

High-frequency noise can lead to enhanced transitions both from the singlet to the triplet state and from the triplet state back to the singlet.   To quantify the net effect,  let us  define  $P_T(v, \beta)$ as the probability to obtain a triplet state in the presence of noise, after a Landau-Zener sweep that starts in the singlet state,  with an off-diagonal matrix element $v$ and a sweep rate $\beta$. Kayanuma~\cite{Kayanuma1984}  has shown that  $P_T$ and $P_{LZ}$ are identical for small $\gamma$, to first order order in $\gamma$, but more generally, $P_T < P_{LZ}$.  
In the limit of strong noise, he obtains an analytic form:
\be
\label{PKay}
P_T \to P_{SN} = \frac{1 - e^{- 4 \pi \gamma}}{2} .
\ee
 
The effects of high frequency noise on the triplet-return correlations should be accounted for, in principle, by replacing $P_{LZ}(\gamma)$ by $P_T(v,\beta)$ in the formulas derived in the previous sections.   In the linear approximation of Subsection III B, where  areas of the Gaussian peaks in $S(\nu)$ were calculated only to  lowest order in $\gamma$, these results would be unaffected by high-frequency noise.  Beyond lowest order, however, we expect that the effects of non-linearity, which tend to reduce the areas of the low-order peaks and increase the amplitudes of higher-order peaks, should be enhanced when $P_{LZ}$ is replaced by $P_T$, assuming that the sweep rate is adjusted to keep the mean value of $P_T$ fixed.  For example, if we consider the (approximate) expression (\ref{flala}), we see that the value of  $f_{\lambda \lambda'}$, should be decreased by the charge noise, as the derivative  $P'_T$ should be smaller than $P'_{LZ}$ , for equal values of the transition probabilities.

We would also expect the magnitude of the  frequency-independent stochastic background term, $N_\tau f_B$, to be increased by high-frequency  charge noise. Specifically, if one uses the large-noise expression (\ref{PKay}) instead of $P_{LZ}$ in the analysis of Subsection III D, one finds
\be
f_B \to \frac{\langle P_T \rangle}{1 +2 \langle P_T \rangle} >  
\frac {  \langle P_T \rangle \,   (1-\langle P_T \rangle ) }   {1+\langle P_T \rangle}  .
\ee
Thus the $f_B$ in the presence of strong noise is larger than the value without noise at equal values of the mean transition probability $\langle P_T \rangle$.
In the limit of large noise and slow sweep rates, where $P_T \to 1/2$, we see that $f_B \to 1/4$, which is its largest possible value.

\subsection{Intermediate-Frequency Noise}

Noise fluctuations may be described as intermediate frequency, if their frequency is low compared to $v$, but comparable to or larger than  the inverse of the time separation $\tau$ between successive LZ sweeps.  
Such fluctuations will have no significant effect on the average spin-flip probability in a single sweep, but they can give rise to fluctuations in the actual times of the 
 $S$-$T_+$ anticrossings.  One possible way of incorporating this in our formalism would be by treating the time between pulses to have a fluctuating part, i.e. $t_p = t_p^0 + \delta t_p$, where $t_p^0= p \tau$ is the equally spaced regular component, while $\delta t_p$ is the fluctuation at pulse $p$.  We assume that there are no correlations in $\delta_p$  from one sweep to the next, and we assume that  $\delta t_p$ has a Gaussian distribution:
\begin{equation}    
p(\delta t_p) = \sqrt{\frac{2}{ \pi \delta_t^2}} \exp \left(-2\left(\frac{\delta t_p}{\delta_t}\right)^2 \right)
\label{eq:p_tp}
\end{equation}

At least within the linear approximation in which we keep only terms of order $\gamma$ in the expansion of $P_{LZ}$,
we can examine the effects of these fluctuations.  
Including the fluctuating part of $t_p$ and $t_q$  in Eq.~(\ref{eq:ChiCorr}), and averaging over the probability distribution in Eq.~(\ref{eq:p_tp}), we find that  the height of the peaks in the power spectrum is weakened. 
Performing the Gaussian average of the terms in Eq.~(\ref{eq:S_woso}),  neglecting the contributions of $\delta t_p$ to the terms originating from relaxation of correlation in the nuclear spin environment and also assuming that the fluctuations at time $t_p$ and $t_q$ are uncorrelated, the effect on the finite frequency peaks due to charge noise can be simply expressed by replacing  the function $G_{\lambda \mu}$ defined in Eq.~(\ref{eq:G_la}) by 
\begin{equation}
G^{{\rm{cn}}}_{\lambda \mu} (\nu) = \exp \left( - \pi^2 (\nu_{\lambda}-\nu_{\mu})^2 \delta^2_t \right) G_{\lambda \mu} (\nu)
\label{eq:G_la_noise}
\end{equation}
Since $\nu_\lambda$  is proportional to the applied field $B$, for any given species, the decrease in peak intensity predicted by (\ref{eq:G_la_noise}) should become more pronounced with increasing $B$. 
We expect that weight lost from the Gaussian peaks will largely reappear in the frequency independent background, but this has not been analyzed in detail. 

According to the current analysis, charge noise at frequencies smaller than the inverse of the total time scale of a run, $N_\tau \tau$, should have no effect on the measured correlations, as it will lead to a uniform shift in crossing times of all sweeps.  We assume here that the  low-frequency noise  is not large enough to cause changes in the electronic wave functions that could affect the value of $P_{LZ}$.

\section{Comparison between theory and experiment}

In this section we examine the extent of  agreement between the various approximations of the theoretical model and the experimental observations\cite{note1}. 

The correlation power-spectrum $F(\nu_n)$,  which is the main quantity of interest to us, was defined in Eq. \ref{Fnun1} and was plotted in Figs.~\ref{fig:MultiSweep_PLZ_wo_SOI} and \ref{fig:MultiSweep_PLZ_with_SOI}. For the experiments under consideration the parameters which have been varied are the rate of the LZ sweep $\beta$, and the direction $\phi$ and magnitude of the magnetic field $B$. For experimental data shown in Figs.~\ref{fig:PLZ_corr_B_0.19_PLZ_0.4} and \ref{fig:PLZ_corr_B_0.4_PLZ_0.4} (from now on referred to as Experiments A and B, respectively), the magnetic field was aligned with the spin-orbit field ($\phi=0$), so we may assume that  spin-orbit coupling $v_{so}= 0$. In a second set of experiments shown in Figs. \ref{fig:PLZ_corr_B_0.1_PLZ_0.6_phi5} (Experiment C) and \ref{fig:PLZ_corr_B_0.19_PLZ_0.42_phi10} (Experiment D), the magnetic field was at a non-zero angle to the spin-orbit field, so $v_{so} \neq 0$. 

A comparison between the principal experimental results and the theoretical predictions described in Sec.~III is summarized in Table I.  For each experiment, A - D, there are four rows in the table, corresponding to the linear approximation of Subsection III B, the nonlinear approximation of  Subsection III C, the full calculation of Subsection III D, and the experimental results.  The column labeled $\langle P_{LZ} \rangle$ shows the theoretical predictions and experimental results for the mean value of the electronic triplet return probability for a single Landau-Zener sweep in each of the four experiments.  Columns labeled $F_{\lambda \lambda'}$ are the areas under the Gaussian peaks in $F(\nu)$ centered at the differences of the Larmor  frequencies  for the indicated nuclear species.    The columns  labeled    $F_{\lambda}$ are the areas under the peaks centered at the Larmor frequencies of individual Larmor species, which are present only for $v_{so} \neq 0$.  The column labeled $\Delta F_0$ is the area under the full peak around $\nu = 0$, excluding the singular $\delta$-function  contribution from the point precisely at $\nu=0$. [Cf. Eq.~(\ref{Fnun2}).]  The column $F_B$ shows the theoretical predictions and experimental results for  the frequency-independent background count.  No values have been entered, in  this column, on  lines corresponding to the linear and nonlinear approximate theories, as the full theory of Sec. IIIA for $F_B$ is already simple.

Theoretical predictions for  the  measured  peak areas  and background counts are related to the intensive quantities calculated in Sec. III by $F_B = N_\tau f_B,$ $  F_{\lambda \lambda'} = (N_\tau/ \tau) f_{\lambda \lambda'} ,$ $ F_{\lambda} = (N_\tau/ \tau) f_{\lambda } ,$ and $\Delta F_{0}  = (N_\tau/ \tau)  (f_0 - f_\infty)$, where $N_\tau = 500$ in these experiments, and $ (N_\tau/ \tau) = $125,000 kHz.  The  input parameters  for these calculations were obtained  from the measurements  reported in  Ref.~\cite{Nichol2015}, which were taken with  high sweep rates, where  $\gamma$ was small enough that the linear approximation is reliable.  The value of $v_{so}$ used for cases C and D were chosen as $v_{so} = \Omega_{so} \sin \theta \sin \phi$,  where $\Omega_{so} = 461 \pm 10$ neV, and $\theta$ is the (11)-(02) mixing angle at the $S-T_+$ crossing point, defined in Ref.~\cite{Nichol2015}, which depends on the strength of the applied magnetic field.  We have extracted values of $\sin \theta$ for the fields used in experiments C and D from the plots in Ref. \cite{Nichol2015}.

\renewcommand{\arraystretch}{1.2}
\begin{widetext}
\begin{center}
\begin{table}[!hbtp]
\begin{tabular}{|c|c|c|c|c|c|c|c|c|c|}
\hline 
Thy./Exp. & $\langle P_{LZ} \rangle$ & $^{69}Ga$ - $^{71}Ga$ & $^{71}Ga$ - $^{75}As$ & $^{75}As$ - $^{69}Ga$  & $\nu \approx 0$ & $^{69}Ga$ & $^{71}Ga$ & $^{75}As$ & $F_B$ \\ 
&    &$F_{\lambda \lambda' }$ (kHz)   &$F_{\lambda \lambda' }$ (kHz)    &  $F_{\lambda \lambda' } $ (kHz) &  $\Delta F_{0 }$ (kHz) &$F_{\lambda  }$ (kHz)  & $F_{\lambda }$ (kHz)  &$F_{\lambda  }$ (kHz)  &\\ \hline
Expt. A & $\mathbf{0.39 (0.05)}$ & $\mathbf{336 (12)}$ & $\mathbf{611 (11)}$ & $\mathbf{623 (12)}$ & $\mathbf{1830 (16)}$ & $0$ & $0$ & $0$ & $\mathbf{98.8 (0.3)}$\\ \cline{1-10}
Full & $\mathbf{0.44 (0.03)}$ & $\mathbf{544 (41)}$ & $\mathbf{1188 (70)}$ & $\mathbf{1109 (66)}$ & $\mathbf{341 \cdot 10^1 (16)}$ & $0$ & $0$ & $0$ & $\mathbf{85.5 (0.6)}$\\ \cline{1-10}
Non-linear & $0.47 (0.04)$ & $648 (40)$ & $1143 (41)$ & $1085 (41)$ & $391 \cdot 10^1 (20)$ & $0$ & $0$ & $0$ & n/a \\ \cline{1-10} 
Linear & $0.79 (0.10)$ & $42 \cdot 10^2 (11) $ & $101 \cdot 10^2 (27)$ & $94 \cdot 10^2 (25)$ & $ 310 \cdot 10^2 (82)$ & $0$ & $0$ & $0$ & n/a \\  \specialrule{.2em}{.1em}{.1em}

Expt. B& $\mathbf{0.40 (0.04)}$ & $\mathbf{248 (12)}$ & $\mathbf{393 (13)}$ & $\mathbf{419 (12)}$ & $\mathbf{1097 (15)}$ & $0$ & $0$ & $0$ & $\mathbf{106.6 (0.3)}$\\ \cline{1-10}
Full & $\mathbf{0.56 (0.03)}$ & $\mathbf{652 (18)}$ & $\mathbf{1343 (15)}$ & $\mathbf{1257 (15)}$ & $\mathbf{3680 (22)}$ & $0$ & $0$ & $0$ & $\mathbf{78.8 (2.5)}$ \\ \cline{1-10}
Non-linear & $0.60 (0.03)$ & $716 (5)$ & $1142 (36)$ & $1092 (32)$ & $4237 (30)$ & $0$ & $0$ & $0$ & n/a \\ \cline{1-10}
Linear & $1.28 (0.14)$ & $109 \cdot 10^2 (24)$ & $264 \cdot 10^2 (59)$ & $246 \cdot 10^2 (55)$ & $81 \cdot 10^3 (18)$ & $0$ & $0$ & $0$ & n/a \\ \specialrule{.2em}{.1em}{.1em}

Expt. C& $\mathbf{0.60 (0.06)}$ & $\mathbf{213 (16)}$ & $\mathbf{306 (15)}$ & $\mathbf{283 (16)}$ & $\mathbf{848 (18)}$ & $\mathbf{76 (14)}$ & $\mathbf{89 (56)}$ & $\mathbf{104 (14)}$ & $\mathbf{106.1 (0.6)}$ \\ \cline{1-10}
Full & $\mathbf{0.69 (0.02)}$ & $\mathbf{597 (14)}$ & $\mathbf{1167 (38)}$ & $\mathbf{1095 (35)}$ & $\mathbf{332 \cdot 10^1 (11)}$ & $\mathbf{129 (8)}$ & $\mathbf{135 (8)}$ & $\mathbf{200 (11)}$ & $\mathbf{62.5 (3.3)}$ \\ \cline{1-10}
Non-linear & $0.75 (0.02)$ & $530 (43)$ & $750 (76)$ & $724 (72)$ & $326 \cdot 10^1 (20)$ & $34.4 (6.0)$ & $35.6 (6.3)$ & $48.6 (9.5)$ & n/a\\ \cline{1-10}
Linear & $2.28 (0.23)$ & $335 \cdot 10^2 (68)$ & $81 \cdot 10^3 (16)$ & $75 \cdot 10^3 (15)$ & $249 \cdot 10^3 (51)$ & $469 \cdot 10^1 (85)$ & $504 \cdot 10^1 (92)$ & $113 \cdot 10^2 (21)$ & n/a \\ \specialrule{.2em}{.1em}{.1em}

Expt. D& $\mathbf{0.42 (0.05)}$ & $\mathbf{124 (23)}$ & $\mathbf{334 (90)}$ & $\mathbf{169 (18)}$ & $\mathbf{595 (23)}$ & $\mathbf{246 (15)}$ & $\mathbf{256 (15)}$ & $\mathbf{468 (18)}$ & $\mathbf{106.8 (0.6)}$\\ \cline{1-10}
Full & $\mathbf{0.42 (0.03)}$ & $\mathbf{274 (18)}$ & $\mathbf{660 (43)}$ & $\mathbf{613 (40)}$ & $\mathbf{263 \cdot 10^1 (22)}$ & $\mathbf{531 (56)}$ & $\mathbf{566 (59)}$ & $\mathbf{1070 (98)}$ & $\mathbf{86.2 (0.2)}$ \\ \cline{1-10}
Non-linear & $0.43 (0.03)$ & $409 (29)$ & $765 (38)$ & $722 (37)$ & $193 \cdot 10^1 (11)$ & $872 (65)$ & $922 (70)$ & $163 \cdot 10^1 (15)$ & n/a \\ \cline{1-10}
Linear & $0.72 (0.09)$ & $228 \cdot 10^1 (56)$ & $55 \cdot 10^2 (14)$ & $51 \cdot 10^2 (13)$ & $169 \cdot 10^2 (42)$ & $44 \cdot 10^2 (10)$ & $48 \cdot 10^2 (11)$ & $107 \cdot 10^2 (25)$ & n/a \\ 
\hline
\end{tabular}
\caption{Comparisons between theoretical calculations and experimental data. Columns 3 to 9 show areas under various peaks in the frequency spectra of the triplet-return correlation functions, obtained from various theoretical estimates or from fits to the experimental data, for four experiments (A, B, C, and D) with different parameters. Column 10, labeled $F_B$ shows the frequency independent background of the correlation  spectrum, while  column 2, labeled $\langle P_{LZ} \rangle$, shows the mean value of the triplet return probability, predicted by the various theories or experimentally observed. The numbers in the parentheses are the estimated errors in the average values. Further explanations are given in the text. The magnetic field values for experiments A, B, C, and D are $0.19$ T, $0.40$ T, $0.10$ T, and $0.19$ T respectively. For experiments A and B the magnetic field is pointing along $\phi=0$, while C and D have $\phi=5^{\circ}$ and $10^{\circ}$ respectively. The experimentally controlled sweep rate $\beta$ in units of $10^{-3} \mu$eV$^2$ are $9.2 \pm 1.1$ (A), $5.67 \pm 0.54$ (B), $3.24 \pm 0.27$ (C), and $12.4 \pm 1.3$ (D). The errors have been rounded off to $2$ significant figures.
}
\label{table:Area}
\end{table}
\end{center}
\end{widetext}

The  rms values  $\sigma $  of the  $x$- and $y$-components of the effective nuclear Overhauser fields  are also expected to depend on the applied magnetic field, and should  be  fit, following Ref.~\cite{Nichol2015},  with a form $2 \sigma ^2 = \sigma^2_{HF} \cos^2 \theta$, where $\sigma_{HF} \approx 34 \pm 1$ nev.  Since $\cos \theta$  is  close to unity in all our experiments, we have ignored the $\theta$-dependence in our calculations and simply used $\sigma = 24$ nev for all four experiments. 

The uncertainty in the parameters $\beta$, $\sigma$, and $\Omega_{so}$ extracted from the experiments lead to errors in the theoretically predicted areas ($F_{\lambda}$, $F_{\lambda \lambda'}$) and the background ($F_B$). In the linear and non-linear approximation, errors in the theoretical values were estimated using the algebraic relations in Sections \ref{subsec:linear} and \ref{subsec:non-linear}. For the full calculation, $150-200$ realizations of the triad of quantities were generated from a Gaussian distribution with widths given by their experimental errors. The error due to a specific parameter was estimated by fixing the rest of the parameters to their average values, and evaluating the quantity of interest for the distribution. This was repeated for each individual parameter and the total error in the predicted value of the quantity was calculated by summing the errors in quadrature. These errors are listed in Table I next to the average values within parentheses. 

We emphasize that effects of charge noise have not been included in the theoretical results presented in Table I.

Experimental values listed in Table I,  and in Table II below, were obtained by fitting the experimental data shown in the figures to the sum of a constant background, a half-Gaussian peak centered at $\nu=0$, and three or six Gaussian peaks centered at finite frequencies, depending on whether $v_{so}=0$, as in experiments A and B, or $v_{so} \neq 0$, as  in C and D.  In each case, the areas and widths of the fitted Gaussians were taken as adjustable parameters, as were the center positions of the finite-frequency peaks. The error in the fitted quantities were evaluated using a least-squares fitting procedure. The resulting fits were shown as the magenta curves in Figs. 2 and 3 and the errors are shown within brackets in Table I and II. 

As explained in Sec III, because the experimental measurements were performed at a series of equally-spaced discrete times, separated by $\tau=4 \mu$sec, the frequencies entering in the measured power spectrum  $F(\nu_n)$ may be restricted to the half Brillouin zone  $0 \leq  \nu_m \leq 1/ (2 \tau) = 125$ kHz, taking into account the inversion symmetry of the underlying power spectrum $S(\nu)$.  [See Eq. (\ref{Fnun3}).]   
Accordingly, an observed peak  in $F(\nu_m)$ whose center frequency $\nu_c$ is closer than its width to the zone boundary 125 kHz, is actually the sum of contributions from two peaks in $S(\nu)$, with center frequencies at $\pm (\nu_c - l/\tau)$, for some integer $l$. In such cases, therefore, in order to compare with the theoretical computations of $S(\nu)$, the experimental areas listed in Table I have been  reduced from the fitted areas by a  factor of two.

As was remarked in Sec. III A, the center positions of the fitted Gaussian peaks agree very well, in all cases, with the known values of the Larmor frequencies of the three species, \cite{Schliemann2003}   or with the differences between them, when aliased  back into the half  Brillouin zone $0 \leq \nu_m \leq 125$ kHz, as predicted by theory. However, as seen from Table I, the areas predicted by the simple linear theory are very much larger than those seen experimentally.  This is not surprising, because the experiments reported here were all performed under conditions where the $P_{LZ}$ is not close to zero, and the linear theory is not adequate.  Results of the non-linear approximation,  shown in the table, are much smaller than the linear approximation results, and are much closer to the experimental results.  Results of the full theoretical calculation are in some cases smaller than those of the nonlinear approximation and in some cases larger; however the full calculations and  the non-linear calculations do not differ by a large factor.  Theoretical predictions for the ratios between peak areas,   within any one of the four  experiments,  are  apparently in reasonable agreement with the experimentally measured ratios.  However, the absolute values of the theoretically predicted areas are still larger than experiment by factors of $~ 2$ or more.
 
We believe that a large part of the remaining discrepancies can be attributed to the effects of charge noise, which have been omitted from the calculations shown in Table I, and which are known to be significant under the conditions of these experiments. In earlier experiments performed by Nichol et al. \cite{Nichol2015} with measurements on the same device  the influence of charge noise was mitigated by employing LZ protocols with fast sweeps. According to Kayanuma \cite{Kayanuma1984}, asymptotic behavior of  the triplet return probability $P_T$ at large $\beta$ is unaffected by the presence of high-frequency white noise. Though in real experiments the noise may be colored, numerical simulations with realistic parameters showed that this behavior still survived beyond  Kayanuma's theoretical  approximation. 

The model parameters  $\sigma$ and ${v}_{so}$  used in our calculations were obtained from analysis of the  behavior of  $\langle P_T \rangle$ at large $\beta$, so they may be considered reliable despite the effects of noise.   Both   $\sigma$ and $v_{so}$  are independent of sweep rate, and they should not be directly affected by noise.  However, for the slower  sweep rates used in the current experiments, the value of  $\langle P_T \rangle$  is, apparently, already affected by noise, according to the data and simulations shown in Fig. 1 of the Supplementary Material for Ref.~\cite{Nichol2015}. Extending the picture incorporated in the  nonlinear approximation developed in subsection III C, we would expect the effect of charge noise on the areas of the correlation peaks to be much greater than  on the mean value  $\langle P_T \rangle$.  As discussed  in Sect. IV above, if  the value of $P_T$ saturates, for large values of $\gamma$ at a value much below the asymptotic value of unity for the case without noise, then the value of $P'_T $,which should actually appear, for example in Eq. (\ref{flala}),  could be much smaller than the value of $P'_{LZ}$ in the absence of noise,  even if the values of $\gamma$ are chosen in the two cases to make the mean value of $P_T$ the same.  Since the square of $P'_T$ enters  in the nonlinear renormalization of the peak area, it  seems quite plausible that high-frequency charge noise could be responsible for much of the remaining discrepancies between theory and experiment.

As discussed in  Sec. IV, effects of high frequency charge noise might also be a reason why the observed background counts $F_B$ are higher, by about 20\%, than  predictions of the theory without noise.  

Besides the areas of peaks in the correlation power spectrum and the size of the frequency-independent background,  the widths of the Gaussians fitted to the experimental data may provide a window into the decoherence of the nuclear spin ensemble. The parameters $\tau_{\lambda \mu}$ and $\tau_{\lambda}$ characterizing the phase-coherence time [see Eq.~(\ref{eq:GaussDecay}) and (\ref{tlmtltm})] extracted from the Gaussian fits are shown in Table~\ref{table:Width}.  It appears that the values of $\tau_{\lambda}$ extracted from the single-frequency peaks in the presence of spin-orbit interactions do not bear the relation to the values of $\tau_{\lambda \mu}$ given by Eq.~(\ref{tlmtltm}), but the scatter in the experimental values of $\tau_{\lambda}$ is large, and it is difficult to attach significance to these numbers. As one source of error, we note that due to non-linear effects, there should be weaker peaks at various combinations of the harmonics of the nuclear Larmor freqencies, which are not generally resolved in the data, and the fitted width of one of the six basic peaks may be erroneously large, if there is an overlap with one or more of these unresolved peaks.

\begin{widetext}
\begin{center}
\begin{table}[!hbtp]
\begin{tabular}{|c|c|c|c|c|c|c|c|}
\hline
Exp. & $^{69}Ga$ - $^{71}Ga$ & $^{71}Ga$ - $^{75}As$ & $^{75}As$ - $^{69}Ga$ & $\nu \approx 0$ & $^{69}Ga$ & $^{71}Ga$ & $^{75}As$\\ \hline
A & $57.7 (1.5)$ & $59.2 (1.6)$ &  $52.3 (1.2)$ & $56.55 (0.57)$ & $\emptyset$ & $\emptyset$ & $\emptyset$ \\ \hline
B & $53.6 (1.3)$ & $51.1 (1.2)$ &  $52.3 (1.2)$ & $57.71 (0.59)$ & $\emptyset$ & $\emptyset$ & $\emptyset$ \\ \hline
C & $52.3 (1.2)$ & $57.7 (1.5)$ &  $53.6 (1.3)$ & $60.83 (0.66)$ & $45.5 (7.8)$ & $69 (30)$ & $45.5 (5.2)$ \\ \hline
D & $52 (11)$ & $62 (12)$ &  $66.2 (7.8)$ & $72.6 (2.3)$ & $58.9 (4.4)$ & $75.8 (3.6)$ & $66.3 (2.8)$ \\ \hline
\end{tabular}
\caption{$\tau_{\lambda \mu}$ and $\tau_{\lambda}$ in $\mu$s extracted from Gaussian fit of spectral peaks in experimental data. The numbers in the parentheses are the estimated errors from the least-squares fit of the experimental data. The errors have been rounded off to $2$ significant figures.}
\label{table:Width}
\end{table}
\end{center}
\end{widetext}

The magnitudes of the observed peak widths are roughly consistent with an 
interpretation that the primary reason for decay of the nuclear spin correlation functions $g_\lambda (t) $ is some combination of electric quadrupole effects and  an inhomogeneous broadening of the nuclear Larmor frequencies 
due to  interactions between neighboring  nuclear spins \cite{Shulman1958,Poggio2005}. Our measured line widths are also approximately consistent with previous measurements of nuclear magnetic resonance line widths at low temperature in GaAs quantum wells\cite{Poggio2005}. 
Also, as discussed in the Appendix, below,  these interactions should lead to  a distribution of the Larmor frequencies for each species that is roughly Gaussian in shape, which would lead to a roughly Gaussian shape for the time-dependent nuclear spin correlations $g_\lambda (t)$, given by Eq.~(\ref{eq:GaussDecay}), as we have assumed in the analysis of subsection III A. However, the deviations from a Gaussian distribution may have an important effect on the triplet return correlations at the lowest frequencies.

\subsubsection{Excess weight at the lowest frequencies.}

In contrast with the peaks centered at finite frequencies, the peak centered at zero frequency is not so well fit by a Gaussian shape.  This is evident in Figure  4, which presents expanded views  near zero frequency of the power spectra of the four experiments shown in  Figures 2 and 3. 
Within the linear  theory of subsection III-A, the peak around zero frequency should be the sum of three  Gaussians of possibly different widths, arising from the different nuclear species, which might account for some of the deviation from simple Gaussian behavior.  The effects of nonlinearities on the line shape have not been explored carefully.  However, there is one striking feature of  the data shown in Fig. 4 that cannot be explained by non-linearity alone -- i.e., the significant amount of extra weight in the intensity at $\nu = 0.5$ kHz, which is the lowest non-zero frequency in our discrete Fourier transform.  In all cases, these intensities are larger, by $~$50 to 100 counts than the value one would expect extrapolating the intensities from the nearby points, at 1-4 kHz.
This sharp feature points to some sort of drift or low-frequency noise, with correlations that exist on time scales of 2~ms, the duration of a run of 500 Landau-Zener sweeps. 

\begin{figure}[!hbtp]
\subfloat[]{
\includegraphics[width=3.0in]{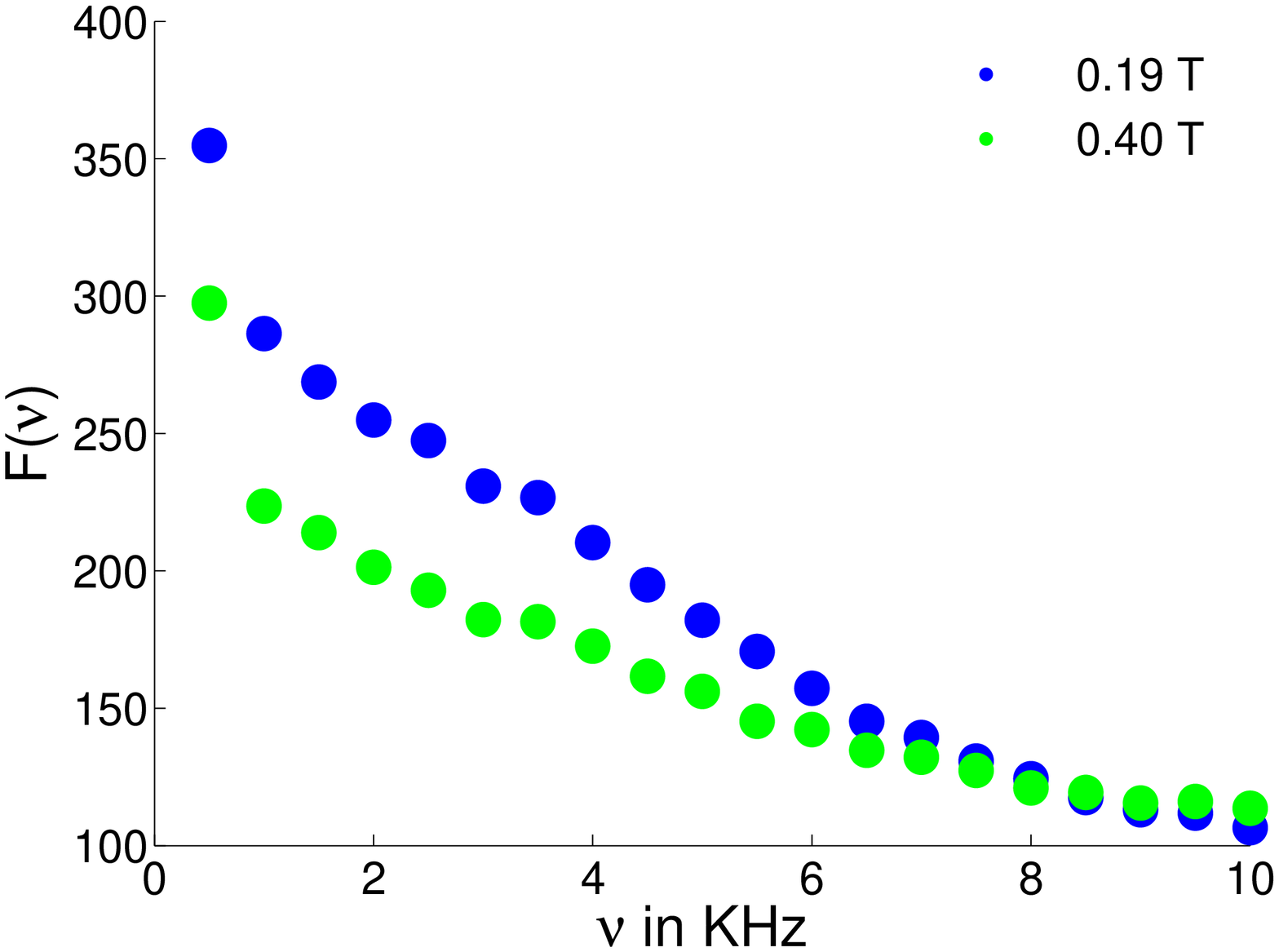}
\label{fig:PLZ_corr_low_freq_phi_0}
}
\\
\subfloat[]{
\includegraphics[width=3.0in]{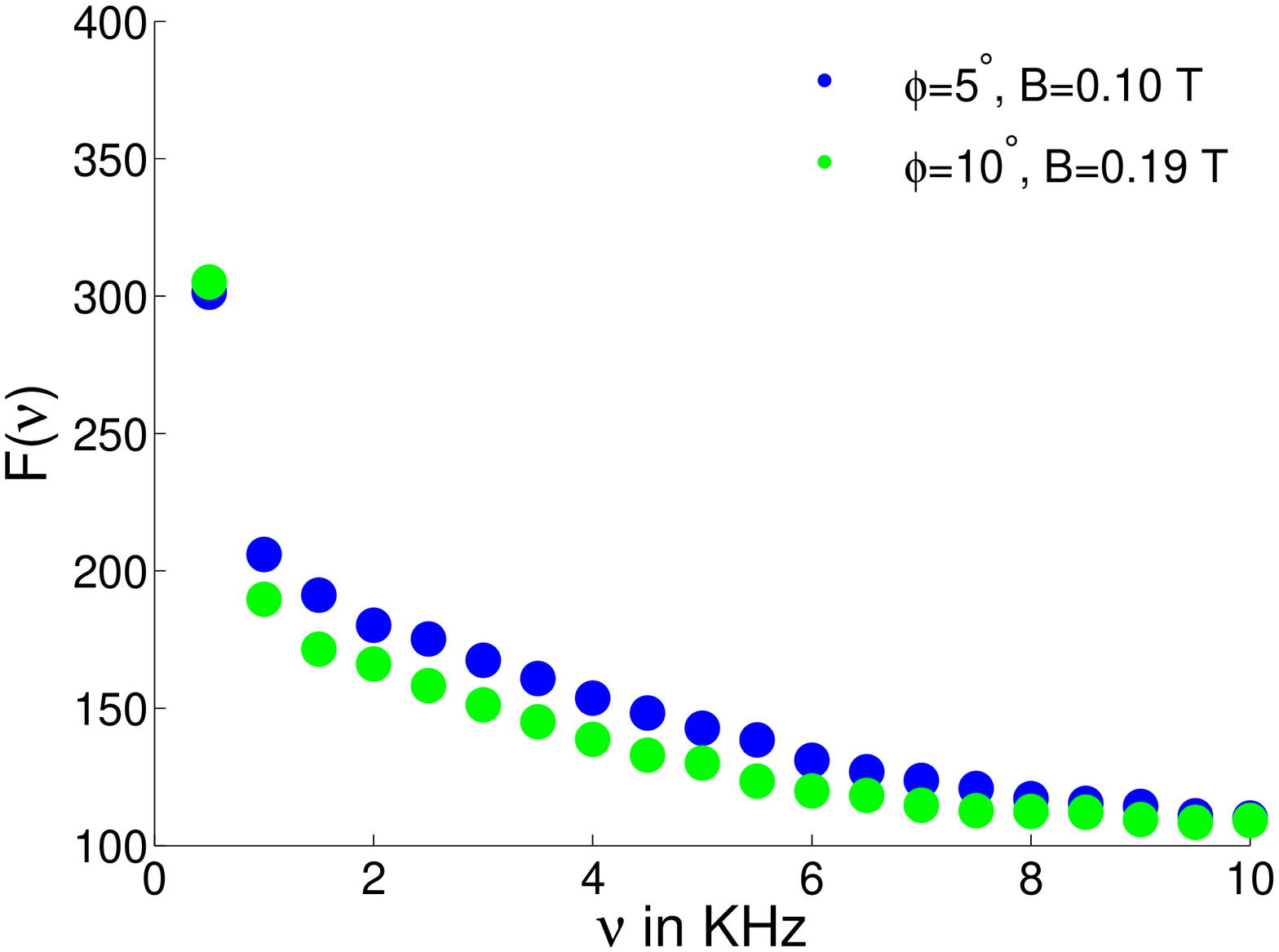}
\label{fig:PLZ_corr_low_freq_phi_5}
}
\caption{The weight at low frequencies in the power spectrum $F(\nu_n)$. Panel (a) is at field orientation $\phi=0$, where spin-orbit effects are absent, with magnetic field strengths $B= 0.19$ T and $0.40$ T. Panel (b) is at field strengths $B=0.10$ T  and $0.19$ T, at orientations $\phi = 5^{\circ}$ and $10^{\circ}$, respectively, where SO is effective. In both cases the excess weight at low frequencies is prominent as non-Gaussianity in the experimental data.}
\label{fig:Excess_weight_low_freq}
\end{figure}

A related anomaly occurs in the fluctuation contribution to the intensity at {\em zero frequency}, which we may define by 
\be
\delta F(0) \equiv \langle {\tilde{\chi}_0^2 }\rangle  -  \langle {\tilde{\chi}_0 }\rangle^2 .
\ee
The experimental values of $\delta F(0)$,  (not shown in the plots), were found to be  
\be
\delta F(0) = 555, \, \, 441, \,\, 934, \,\, 516 ,
\ee
 for Experiments A-D, respectively. 
These values are again larger, by amounts ranging from $~$200 to 700 counts, than the numbers one would obtain  by smoothly extrapolating the values of $F(\nu_m)$ from small non-zero $\nu_m$ to $\nu=0$, which are the values  one would have expected to find for $\delta F(0)$  in the absence of drift or low-frequency fluctuations. 

At present, we do not have a clear explanation for the extra intensity at our lowest frequencies.  We have checked for a  possible systematic drift in the triplet return probability during the course of 500 sweeps by separately calculating the  triplet return probability averaged over sweeps in the first, second, third, and fourth groups of 125 sweeps within a run, for each of our  four experiments, averaged over the 14,400 runs accumulated for each experiment.   The data suggests the possibility of a small downward drift in the triplet return probability by an amount of the order of a fraction of one percent, but this amount is comparable to the fluctuations in the data, and may not be statistically significant. In any case, a drift of this amount is far too small to explain the observed extra weight at the lowest frequencies.  

A possible origin of the low-frequency anomaly in the triplet return correlation spectrum may arise from residual long-term correlations of the nuclear spins. This possibility is discussed in the Appendix, below.

\section{Conclusions}

 In this paper, we have presented experimental data and associated theory for correlations in the triplet return probability in a series of  experiments involving repeated Landau-Zener sweeps through the crossing point of a singlet state and a spin aligned triplet state in a double quantum dot (DQD) containing two conduction electrons.
As the probability of an electron spin flip is strongly influenced by the nuclear Overhauser fields transverse to the applied magnetic field, correlations in the triplet return probability are sensitive to  correlations in the nuclear orientations.  The  experiments reported here employ a series of 500 sweeps that are separated by  intervals  $\tau = 4 \,  \mu$s, so they measure correlations on time scales from 4 $\mu$s to 2 ms.

Our theoretical analysis employs a semi-classical description of the  transverse nuclear spin components. Neglecting complications such as the effect of  high-frequency charge noise during a Landau-Zener sweep, the probability $P_T$ of triplet-return in a given Landau-Zener sweep should have the Landau-Zener form,
$P_{LZ} = 1 - e^{- 2 \pi \gamma}$, where $\gamma$ is proportional  to the absolute square of the sum of the transverse nuclear Overhauser fields and the spin-orbit field and is inversely proportional to the Landau-Zener sweep rate.  The effective spin-orbit field depends on the strength and direction of the in-plane magnetic field,  and it may be eliminated if the magnetic field is aligned in a direction determined by the orientation of the axis of the DQD. 

 Correlations in the triplet return probability are most conveniently discussed in terms of the frequency-dependent power spectrum $F(\nu)$.  In the cases where the spin-orbit field is absent, the most prominent  features of  the experimentally measured $F(\nu)$ are a set of peaks centered at $\nu=0$ and at the differences of the Larmor  frequencies of the nuclei, which sit on top of a frequency-independent background.   When the spin-orbit field is non-zero, there are additional peaks, centered at Larmor frequencies of the individual species.  (All frequencies in $F(\nu)$  should be interpreted modulo  $1/ \tau$, due to the periodic spacing of the sweeps.)  
 
Our  theoretical analysis correctly predicts the positions of the observed peaks, and gives a reasonably accurate prediction of the size of the frequency-independent background.  However, a theoretical analysis neglecting  the effects of high-frequency charge noise predicts peak areas that are larger than the observed areas by a factor of two or more. Our estimates suggest that the effects of high-frequency charge noise may be responsible for this discrepancy, but we have not attempted a quantitative calculation of these effects.  The observed peak widths are roughly consistent with  theoretical predictions, which relate these widths to the widths of the  nuclear NMR lines, which might result from inhomogeneous broadening  or other mechanisms, corresponding to time scales for nuclear dephasing of the order of 60 $\mu$s.  However, there is some uncertainty in the fitted experimental peak widths, and we are not able to assert a quantitative understanding of the peak widths. 

In our discussions of the theoretical  predictions for the areas of the principal peaks in $F(\nu)$,  we presented, in addition to the full theory of Section III D, two approximate calculations,  designed to elucidate  the underlying physics. In Section III B, we discussed a linearized theory,  where the exact formula for $P_{LZ}$  was replaced by its linear approximation $2 \pi \gamma$ .  While this approximation should be adequate at sufficiently high sweep rates, where $\gamma$ is small, the approximation fails seriously for the parameter values in our experiments, where the mean values of $P_{LZ}$ are $\ge 0.4$. For the parameters appropriate to our experiments, we find that peak areas obtained from the linearized theory are larger than those predicted by the full theory by factors of  six or more.
In Section III C, we presented an approximate theory that takes into account the  most important effects of the nonlinear dependence of $P_{LZ}$, and which gives predictions that are in reasonable agreement with those of the full theory, as detailed in Table I. 

The experimental values for $F(\nu)$ show excess weight at our two lowest frequencies, $\nu=0$ and $\nu=500$ Hz, which cannot be explained by our theoretical model, with or without the effects of high-frequency charge noise, if we assume Gaussian line shapes for the decay of nuclear spin correlations.  
However, part or all of this excess weight might be explained by deviations from a Gaussian line shape. In particular, if one takes into account broadening due to interactions of the nuclear quadrupole moments with gradients in the local electric field, which can vary from site to site, and if one can neglect all other broadening mechanisms, then the frequency spectrum for the transverse spin correlations of a given nuclear species will consist of a $\delta$-function at the unshifted Larmor frequency, in addition to a component that is  broadened by the quadrupole coupling.\cite{Poggio2005,Botzem2015}   Similarly, if the nuclear spin correlation functions are inhomogeneously broadened due to interactions with nearest neighbor nuclear spins, there could be narrow components at the unshifted Larmor frequencies of the various species that could lead to excess weight at low frequencies in the values of $F(\nu)$.  These possibilities are discussed further in the Appendix below.

In order to clarify further the  sources of extra weight at low frequencies, it would be desirable to conduct additional experiments, with sweep sequences that last longer than the 2 ms used here.   In order to clarify the reasons for deviations between theory and experimental measurements of the areas of the peaks in $F(\nu)$ it would be desirable to do experiments at faster sweep rates, where effects of charge noise should be less important. 

\section*{Acknowledgments}
The authors are grateful for helpful conversations with Arne Brataas and Trevor Rhone. This research was funded by the United States Department of Defense, the Office of the Director of National Intelligence, Intelligence Advanced Research Projects Activity, and the Army Research Office grant W911NF-15-1-0203. S.P.H was supported by the Department of Defense through the National Defense Science Engineering Graduate Fellowship Program. This work was performed in part at the Harvard University Center for Nanoscale Systems (CNS), a member of the National Nanotechnology Infrastructure Network (NNIN), which is supported by the National Science Foundation under NSF award No. ECS0335765.
The work of A.P. was performed in part at the Aspen Center for Physics, which is supported by National Science Foundation grant PHY-1066293.

\section*{Appendix. Quadrupole and  inhomogeneous broadening of the nuclear Larmor frequencies and their consequences for correlation  experiments. }

In the discussions of Section III, we assumed a phenomenological Gaussian form for the correlation function 
$\langle \Omega_{\lambda} (t) \Omega^*_{\lambda} (t') \rangle$ of the transverse hyperfine field for a given species $\lambda$. Here, we discuss two possible mechanisms that might lead to such a frequency broadening of the NMR lines: nuclear quadrupole shifts and inhomogeneous broadening due either to nuclear dipole-dipole coupling of superexchange. Because the resolution of our data is not sufficient to indicate which mechanism is the most important, we consider both effects in detail here.

If a nucleus with spin 3/2 sits in a position with a nonzero electric field gradients, the correlation function for the spin components perpendicular to an applied magnetic field in the z-direction  will be split into three lines \cite{Poggio2005,Botzem2015}.  The portion corresponding to transitions between spin states  $I_z = 1/2$ and $I_z = 3/2$  and that corresponding to transitions between $I_z=-3/2$ and $I_z=-1/2$ will generally be shifted, in opposite directions, by the quadrupole coupling, while the portion corresponding to transitions  between the states $I_z = \pm 1/2$ will be unshifted. The size of the shifts will be proportional to  the magnitude of electric field gradients but will also depend on the orientation of the magnetic field relative to the gradients.  For nuclei in GaAs, the electric field gradients are expected to be proportional to  the local electric field in the vicinity of the nuclear location, and so will have values that vary from one position to another over the thickness  of the electronic wave function.  Therefore, if there is no other mechanism for broadening, the space averaged spectrum for spin fluctuations transverse to the magnetic field will be the sum of  a $\delta$-function at the unshifted Larmor frequency and a  broadened peak whose width is determined by the size of the quadrupole splitting.   The $\delta$-function contribution  to the spectrum, in this case,  would clearly provide a possible explanation for our experimental observations of extra weight in $F(\nu_m)$ at the lowest frequencies.  However, it is less clear how well our observations of apparently Gaussian line shapes for the finite-frequency peaks in  $F(\nu_m)$ can be reconciled with the assumption of purely quadrupolar broadening. 

In a recent experiment, Botzem et al.\cite{Botzem2015}  have investigated nuclear spin correlations using a technique based on electron spin-echo measurements in a  GaAs DQD, and have interpreted the results in terms of a distribution of quadrupole splittings for the  three nuclear species.  For a magnetic field either parallel or perpendicular to the axis of the DQD they report a line width $\delta  B$ which is of order 2 mT for the As nuclei, and is of order 0.4 - 0.5 mT for the two Ga species.  Taking into account the $g$-factors for the different species, this would translate, in our notation, to a quasi-Gaussian (rms) decay time $\tau_\lambda$ of order 30 $\mu$s for  $^{75}$As, and order 60 and 100 $\mu$s for $^{71}$Ga and  $^{69}$Ga.      The value $\tau_\lambda \approx 30 \mu$s for  $^{75}$As is smaller, by a factor of two, than the value  obtained from our Gaussian fit to the data for $F(\nu)$, as listed in Table 2 above,  but this could be due to differences in the quadrupole splittings for the different samples. Thus, it seems plausible that electric quadrupole splitting is the dominant factor in the spectral line width for $^{75}$As.  However, if electric quadrupole splitting were also the dominant factor for the Ga line widths, then the results of Ref.~\cite{Botzem2015} would require that the decay times for the Ga species would be two to three times longer than for $^{75}$As, which is not in accord with our observations. This would then suggest that another mechanism should be responsible for the decay of correlations in the Ga species. 

Nuclear spin correlations were also investigated in Ref.~\cite{Nichol2015} using a protocol in which the triplet return probability in a DQD  was measured after a pair of Landau-Zener sweeps, without reloading the electron. As may be seen in Fig. 3 of that reference, the spectral function for $^{75}$As consisted of a central peak, surrounded by two smaller side peaks, when the field orientation angle was close to 90$^{\rm o}$. The observed splitting between the central peak and the side peaks,  of the order of 10 kHz, is roughly consistent with the estimates of Ref.~\cite{Botzem2015} for the quadrupole effect.  However, the central peak itself appears to have a width larger than the experimental resolution, suggesting that some broadening mechanism in addition to the quadrupole splitting is in play, even for the $^{75}$As.

A second mechanism for NMR broadening is inhomogeneous broadening, where each nuclear spin experiences an effective magnetic field arising from its interactions with the z-component of the nuclear spins on nearby sites. Previous investigations suggest that  dipole-dipole interactions  and superexchange interactions give comparable contributions to the frequency broadening of the NMR lines.\cite{Shulman1958, Sundfors1969} 

If one assumes that the frequency shift of any given nucleus is the sum of contributions from a large number of neighboring nuclei, then one recovers immediately a Gaussian  distribution for the individual frequencies and consequently a Gaussian time-dependence of the correlation function $g_\lambda$  defined in Eq.(\ref{glambda}).  
At the other extreme, however, we may consider a model where only interactions between nearest neighbor nuclei are important.  (This might be a good approximation for the superexchange contribution, but less so for the dipolar interaction.)  If we consider, as an example,  a $^{69}$Ga nucleus  on a lattice site $j$, then the shift of its Larmor frequency will be given by  
\be
\label{dnuj}
\delta \nu_j =  J_{69}  \sum_k I^z_k
\ee
where the sum is over the four $^{75}$As nuclei that are nearest neighbors to $j$, and $J_{69}$ is the appropriate  coupling constant.  Since the As nuclei have spin $I=3/2$, the sum in (\ref{dnuj}) can take on any integer value between -6 and 6, with a maximum probability at $\delta  \nu_j =0$.  The correlation function $g_\lambda$ for $^{69}$Ga can then be written as
\be
\label{G69}
g_{69}(t) = \sum_{n=-6}^{6} w_n \exp (-2 \pi i n J_{69} t \,)
\ee
where $w_n=w_{-n}$ is $2^{-8}$ times the number of ways one can choose a sequence of four integers from the set (-3, -1, 1, 3) so that their sum is equal to $2n$.  As $|n|$ varies from 0 to 6, the quantity  $2^8 w_n$ takes on the values 44, 40, 31, 20, 10, 4, and 1. 

\begin{figure}[!hbtp]
\includegraphics[width=3.0in]{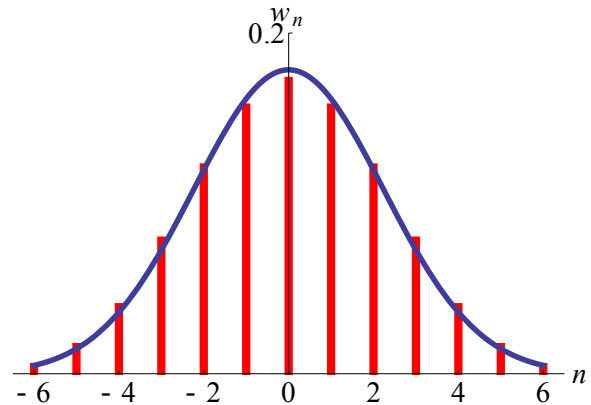}
\caption{Distribution of shifts in the Larmor frequency of a $^{69}$Ga or $^{71}$Ga nucleus due to interactions with the four nearest neighbor $^{75}$As nuclei, leading to an approximately Gaussian decay of the autocorrelation function.  The red vertical lines are the normalized probability of a frequency shift  $nJ_{69}$ or $nJ_{71}$, while the blue curve is a Gaussian fit.}
\label{fig:Freq_dist}
\end{figure}

The values of $w_n$ are compared to a Gaussian with variance 
$\langle n^2  \rangle= 20$, in Figure 5.  The Fourier transform of $g_{69}(t) $
will consist of 13 delta-function peaks at frequencies $n J_{69}$, with weights equal to $w_n$.

The correlation function for $^{71}$Ga should have an identical form to (\ref{G69}) but with a different coupling constant $J_{71}$ instead of $J_{69}$.   The correlation function for $^{75}$As is more complicated because each of its four nearest neighbors can be either of the two isotopes of Ga.  Thus, its Fourier transform will have many more peaks, and its envelope should be even closer to a Gaussian.   
 
Generalizing the arguments given in Section III above, we expect that the line shape for the interference peak near the frequency difference $\nu_\lambda-\nu_{\mu}$ should be proportional to the Fourier transform of the product $g_{\lambda}(t) g_{\mu}(t)$.  For the case $\lambda = ^{69}$Ga, 
$\mu = ^{71}$Ga,  if $J_{69}$ and $J_{71}$ are incommensurate, the Fourier transform will be a sum of  169  $\delta$-function contributions.  If one of the two species is $^{75}$As, the number of distinct $\delta$-functions will be even larger. In either case, when viewed with less than perfect resolution, the line shape should be quite close to the Gaussian form $G_{\lambda \mu}(\nu)$ given in Eq (\ref{eq:G_la}). Second neighbor nuclear interactions, which we have thus far ignored, will further split each $\delta$-function into multiple peaks, which should make the line shapes even more Gaussian-like when viewed with finite frequency resolution.
 We remark that the nonlinear corrections included in Section III C   should have little effect on the line shape of the peak, even though they may greatly reduce the overall area of the peak.  

In contrast, the deviations  of $g_\lambda(t) $ from the Gaussian form (\ref{eq:GaussDecay}) may play a larger role in case of the peak centered at $\nu=0$ in the correlation function $S(\nu)$.   Here we find a contribution from each $\lambda$ that is proportional to the Fourier transform of $|g_\lambda (t)|^2$.  In the case where  $\lambda$  represents $^{69}$Ga or 
$^{71}$Ga, the Fourier transform has only 13 $\delta$-function peaks, so the deviations from a continuous Gaussian may be more pronounced. The most pronounced effect should occur for the $\delta$-function precisely at $\nu=0$, where one expects significant contributions from both Ga species.   The fractional weight of this $\delta$-function, relative to the contribution of the two Ga nuclei to the total area of the peak near zero frequency, should be given by $\sum_n w_n^2  \approx$ 0.13.

When interactions with second and further neighbors are taken into account, the predicted zero-frequency $\delta$-function will split into multiple contributions, slightly shifted from $\nu=0$, so that the peak would effectively be  slightly broadened.  It is possible that such a broadened peak might account for part or all of the extra weight observed experimentally in the correlation functions at our lowest frequencies ($\nu=0$ and 0.5 kHz), as discussed above. 

With regard to quantitative  comparisons between theory and experiment, we note that when correlations are present  on time scales comparable to the experiment duration $N_\tau \tau$, predictions for the observed discrete power spectrum $F(\nu_n)$ should  be extracted from  the predicted correlation function $f(t)$ by replacing $C_\chi(p,q)$  by $f_B \delta_{pq} + f(t_p-t_q)$ in Eq. (\ref{Fnun1}). Then, replacing the summation variable $p$  by $s=p-q$,  the double sum can be reduced to a single sum, with the result  
\be
F(\nu_n) = N_\tau f_B + \sum _{s= -N_\tau }^{N_\tau } e^{2 \pi i n s / N_\tau} f (s \tau) \,
|N_\tau-s| .
\ee

We note that a combination of quadrupole splitting and  inhomogeneous broadening due to interactions between nearest-neighbor nuclei would still lead to a finite  $\delta$-function peak at the unshifted Larmor frequency in the spin  autocorrelation function for each nuclear species, which would still lead to a singular peak at zero frequency in the electronic  triplet-return  spectrum $F(\nu)$.   Whether such a contribution could be large enough to explain our observations remains to be seen.

In addition to quadrupole and inhomogeneous broadening, there may be additional processes which lead to decay of the correlation functions $g_\lambda(t)$, and such processes could also lead to deviations from Gaussian behavior.  For example  flip-flop terms between the nuclear spins would more likely be a Poisson process giving rise to an exponential decay of the correlations, and a  Lorentzian behavior in the Fourier transform.  However,  we expect that the rate for flip-flop transitions would be relatively slow on the time scale of interest here, so this process should not be of importance here. In any case, analysis of our experiments seems to favor the Gaussian assumption.  We also note that Lorentzian behavior would primarily affect the Fourier transform at large frequency shifts, and would not lead to an extra contribution at the smallest frequencies.

\bibliography{References}{}

\begin{thebibliography}{48}
\bibitem[{not()}]{note1}
\bibinfo{note}{We show the comparison between theory and experiment for 4
  representative data sets. We have also collected similar data sets for
  approximately 60 different configurations of magnetic field strengths and
  orientations. In all cases, the data show qualitatively similar results, and
  we can identify the main peaks in the correlation spectra as either
  difference frequencies or bare Larmor frequencies. In some cases, there were
  smaller additional peaks that could be described as side bands to the main
  peaks or might possibly be higher harmonic peaks. Also, in some cases, low
  order peaks were missing from the data, or were too weak to to be clearly
  identified. We have not performed a detailed analysis of the peak areas and
  peak heights in these data sets nor attempted a comparison between theory and
  experiment.}
\end{thebibliography}

\end{document}